\documentclass[fontsize=12pt, oneside, paper=a4, pagesize=auto, titlepage=yes, bibliography		= totocnumbered, listof					= totocnumbered, parskip					= half, abstract				= yes]{scrartcl}

	\pdfoutput=1

	\newcommand{\cpfversion}{Version 1.0.0}
	\newcommand{\cpfbuild}{20211023}

	\usepackage[utf8]{inputenc}
	\usepackage[T1]{fontenc}
	\usepackage[english]{babel}
	
	\usepackage{times}
	\usepackage{eurosym}
	\addtokomafont{descriptionlabel}{\rmfamily}
	\addtokomafont{pagehead}{\scshape}

	\usepackage[activate={true,nocompatibility},final,tracking=false,kerning=true,spacing=true,factor=1100,stretch=10,shrink=10]{microtype}
	
	\usepackage{enumerate}
				
	\usepackage{csquotes}

\usepackage[backend=bibtex]{biblatex}   
\bibliography{papers}

	\usepackage[nottoc]{tocbibind}

	\usepackage[headsepline]{scrlayer-scrpage}
	\ohead{Fries, Christian P.}
	\ihead{Non-Linear Discounting}
	\ifoot{\tiny \copyright 2021 Christian Fries}
	\ofoot{\tiny \cpfversion\ (\cpfbuild )\\ \href{http://www.christianfries.com/finmath}{http://www.christianfries.com/finmath}}

	\usepackage[intlimits]{amsmath}
	\usepackage{amssymb}

	\usepackage{algorithmic,algorithm,listings}
	
	\usepackage{graphicx,xcolor,subfig}
	
	\definecolor{DarkBlue}{rgb}{0.0, 0.0, 0.5}
	\definecolor{DarkRed}{rgb}{0.5, 0.0, 0.0}
	\definecolor{DarkGreen}{rgb}{0.0, 0.5, 0.0}
	\definecolor{DarkYellow}{rgb}{0.5, 0.5, 0.0}
	\definecolor{Brown}{cmyk}{0.00,0.80,1.00,0.60}
	\definecolor{DarkGreen}{cmyk}{0.64,0.00,0.95,0.60}
	\definecolor{DarkBlue}{cmyk}{0.70,0.60,0.00,0.60}

	\usepackage{ifpdf}	
						
	\ifpdf
		\usepackage{hyperref}
		\hypersetup{
			hyperindex = true,
			pdfpagelayout=TwoPageRight,
			pdfborderstyle={/S/U/W 1},
			colorlinks = true,
			linktocpage = true, linkcolor=blue, citecolor=blue, urlcolor=blue, bookmarks,
			pdfauthor={Christian Fries},
			pdfsubject={Interest Rates in Valuation},
			pdftitle={Non-Linear Discounting and Default Compensation},
			pdfstartview=FitH
		}
	\else
		\usepackage{url}
	\fi

	\newcounter{cpf_counter} 
	\newcounter{cpfNumberOfFigures} 
	\newcounter{cpfNumberOfTables} 

\newcommand{\Rset}{\mathbb{R}}

\newcommand{\indicatorfcn}{\mathrm{\mathbf{1}}}
\newcommand{\textsmall}[1]{{\small #1}}

\begin{document}
	
		\subject{}
		\author{
			Christian P.~Fries
			\thanks{Department of Mathematics, University of Munich, Theresienstra\ss{}e 39, 80333 München, Germany}
			\thanks{DZ Bank AG Deutsche Zentral-Genossenschaftsbank, Platz der Republik, 60325 Frankfurt am Main, Germany}
			\thanks{\url{http://www.christianfries.com}}
		}
		\title{
			Non-Linear Discounting \\ and Default Compensation
		}
		\subtitle{
			Valuation of Non-Replicable Value and Damage:\\ When the Social Discount Rate may Become Negative\\[2ex]
			 \textsmall{\cpfversion}\\[2ex]
		}
		\date{June 12th, 2020 \\ {\large(1st update: May 31st, 2021, 2nd update: October 16th, 2021)}}

	\maketitle

	\begin{abstract}
		In this paper, we introduce a model that adds a non-linearity to discounting: the discounting factor may depend on the notional (i.e., discounted values are no longer linear in the notional).

		In the first part of the paper, we provide a discounting when discount factors cannot be derived from market products. That is, a risk-neutralizing trading strategy cannot be performed.
		This is the case when one needs a risk-free (default-free) discounting, but default protection on funding providers is not traded. For this case, we derive a default compensation factor ($\exp(+\tilde{\lambda} T)$) that describes the present value of a strategy to compensate for default (like buying default protection would do).

		In a second part of the paper, we introduce a model where the survival probability, and hence the discount factor, depends on the notional. This model introduces an effect not present in the classical modelling of a time-dependent survival probability. Our model allows that large liquidity requirements are more likely to default instantly than small ones.

		Combined, the two models build a framework where discounting (and hence valuation) is non-linear: discount factors depend on the amount to be discounted.

		The non-linear discounting presented here has several effects, which are relevant in various applications:
		\begin{itemize}
			\item If we consider the question of default-free valuation, i.e., factoring in the cost of default protection, the framework can lead to over-proportional higher values (or cost) for large projects or damages. The framework can lead to the effect that discount factors for large liquidity requirements or projects are an increasing function of time. It may even lead to discount factors larger than one. This effect may have relevance in the assessment of events like those induced by climate change.

			\item For the valuation of defaultable products, e.g., like a defaultable swap, the framework leads to the generation of a continuum of (defaultable) par rate curves (interest rate curves) and the valuation of a payer and a receiver swap differs by more than just a sign.
		\end{itemize}

		Our approach builds on top of the classical theory of discounting (which may either be given as market-implied or be derived from a model of utility, consumption and production). In that sense, it is rather a generalization than an alternative.

		The modelling approach has specific relevance for climate models, where discounting is an important aspect in assessing the severity of future events. Our model may result in non-decaying discount factors (negative discount rates) for certain scenarios.

		Another application would be assessing future costs related to a global event like a pandemic, where costs are on a very large scale. Such large amounts will likely result in a notional dependent discount factor. In addition, one may not default on long term costs related to long term damages, requiring a risk-free discounting of large notional.
	\end{abstract}
	
	\microtypesetup{protrusion=false}
	\begin{footnotesize}
		\vspace*{-10ex}
		\tableofcontents
	\end{footnotesize}
	\microtypesetup{protrusion=true}
	
	\clearpage

	\section{Introduction}

	The concept of valuation tries to determine an equivalent present value for future values. Here, the \textit{value} can be positive (a claim) or negative (a liability or damage).
	Apart from the fact that future values may be uncertain, which may require a concept for risk and a price (or value) assigned to risk, it is required to define the dependency on time. The process of determining the time-value is usually called discounting.
	
	\smallskip
	
	Discount rates or discount factors may be modelled or derived via different approaches. The neo-classical consumption model derives them from consumption, production and utility, see Section~\ref{sec:discountingDamage:economicsConsumptionApproach} below. Alternatively, we may directly infer discount rates from a liquid financial market. A no-arbitrage argument allows relating the two approaches.

	Our approach, however, builds on top of the existence of a financial market providing long-term funding (funding providers) at given rates. In this sense, our approach is built on top of the classical frameworks. Through it, we derive an \textit{effective} (social) discount rate that is different than that obtained from the classical frameworks.
	
	\bigskip

	\paragraph{Market-Implied Valuation}

	In mathematical finance, an approach to derive a value for a financial product is to define it through its current market price. If a financial product does not have a market price, one may try to associate the value of this product with a function of other market observed products by establishing a relation among these products, e.g.~a \textit{replication strategy}.

	\smallskip

	Under suitable (and fairly strong) assumptions, a mathematical theory is then applicable that represents the present value as an expectation of (the distribution of) future values under a (stochastic) model, parametrized solely by market observables. This approach constitutes a \textit{market-implied valuation} with a model using market-implied parameters.
	
	\smallskip

	Market implied valuation is a reasonable approach in many situations, but maybe not in all.  The most critical assumption in this approach is the ability to perform a replication (hedging). With that regard, it should be stressed that even if replication could be performed in theory, a market-implied valuation is not admissible if such a risk-neutralizing replication is not performed in practice.

	\medskip

	We will often use the word value in the following, regardless of whether it is a cost or a benefit, since that is just a matter of the sign. Furthermore, the sign of the value depends on the observer: in a bilateral contract, one counterparty's claim is the other counterparty's liability.

	\subsection{Valuation of a Liability contains the Option to Default}
	\label{sec:discountingDamage:counterInuitiveEffectOfOptionToDefault}

	The market-implied valuation of a liability gives rise to a possibly counter-intuitive dependency on the market-implied default probability. Consider a loan, where a counterparty borrows  a unit $1$ in time $t$, to be paid back in $T$. To compensate for interest and the risk to default, the amount paid back is $1 \cdot \exp(r^{f} (T-t))$ with some rate $r^{f}$. If the counterparty has a larger probability to fail on paying back, the rate $r^{f}$ will be larger.

	Consider a liability where a counterparty is liable to pay unit $1$ at a future time $T > t$. This amount is just a fraction, namely $\exp(-r^{f} (T-t)$, of the payment in the previously mentioned loan. Hence, the value of this liability in time $t$ is that fraction of the corresponding loan taken in $t$, namely $1 \cdot \exp(-r^{f} (T-t))$.

	Now, if the counterparty's creditworthiness decreases, the rate $r^{f}$ will increase (the compensation contracted on loan will increase), and hence, the value of any existing liability will decrease.
	This effect is reasonable from the lenders' perspective since the probability that the borrower defaults on the payment increases. However, the effect appears awkward from the borrowers perspective. Since the value of a liability is negative for him, he profits from an increase in the probability of default. In other words: the borrow sees value in the option to default on its liabilities.

	In a balance sheet, this effect is usually separated as the DVA (debit valuation adjustment).

	\subsection{Valuation of a Damage}

	While the valuation of a liability in the previously discussed form is well-grounded, it cannot be applied to assess the present (time $t$) value of a future \textit{damage} that occurs in time $T$. Consider some environmental damage that needs to be repaired or compensated by all means. Assume some model predicts that the time $T$ value (i.e., cost) of this damage is $V(T)$. It seems tempting to consider the time $t$ value as discounted $V(t) = V(T) \exp(- r^{f} (T-t))$. This may appear reasonable since it is the value that has to be contracted in $t$ to achieve a corresponding payment in $T$.

	Note, however, that such a discounting includes the possibility to default on the liability. However, for the damage, there is no option to default on it.\footnote{With the possible exception to factor in the option of our own extinction.} For that reason, one may conclude that the right way of discounting in this case, would be to use some (idealized) \textit{risk-free} rate $r$ (lower than $r^{f}$) such that $V(t) = V(T) \exp(- r (T-t))$. But then, this approach depends on the ability to perform a risk-free replication, which is - if at all - possible only for liquid market assets.

	Furthermore, the existence of a risk-free interest rate is an illusion, or at best, an approximation only valid for \textit{very} short maturities.

	\medskip

	In this note, we consider the discounting for values (cost) of events that cannot be replicated but have to be compensated by all means. Examples are damages evaluated in climate models or economic damages by pandemics.
	Our framework will consider two (somewhat) independent parts:
	\begin{itemize}
		\item a discounting based on a diversification of default risk, which may lead to discount factors larger than $1$, exhibiting the impact of a possible mismatch of market-implied and realized default probabilities, and,

		\item a valuation, where the (realized) default probability depends on the requested notional, i.e., is non-linear in the notional.
	\end{itemize}
	We will combine the two aspects. Since the market-implied default intensity is a market expectation for common cash-flows and the realized default intensity is state-dependent, large (or huge) cash-flows may receive discount factors larger than $1$.

	\paragraph{Non-Linear Discounting}

	The second part of our framework implies that the discount factor depends on the notional $N$. So instead of a linear map $N \mapsto N \cdot \mathrm{df}(T)$ we will end up with a non-linear map $N \mapsto N \cdot \mathrm{df}(T,N)$. Hence, the framework may be considered as a \textit{non-linear discounting}.

	\subsection{Related Topics}
	\label{sec:discountingDamage:relatedTopics}

	\subsubsection{Time Preference}
	\label{sec:discountingDamage:economicsConsumptionApproach}

	The concept of discounting is rooted in the discounted-utility model dating back to Paul Samuelson in 1937. Starting from this model a discount factor is formed from multiple individual agents, each endowed with an utility function. Then, the discount rate can be derived from the modelling of the utility function in combination with a model for consumption and production. For a broad review of the modelling of ``time-value''  and ``time-preference'' starting with from a discounted-utility model see~\cite{FrederickLoewensteinDonoghue2002Discounting}. A presentation that also relates to the application in climate models can be found in~\cite{GollierWeitzman2010, Gollier2012}. For an overview and relation to alternative methods in (social) discounting, see \cite{HepburnGosnell2014} and references therein.

	We give a short sketch of this framework and its relation to our approach.

	\subsubsection{Deriving the Discount Rate from Consumption Time Preference, Production and Utility}

	As a first component, consider a consumption function $t \mapsto c(t)$ that describes an amount consumed in time $t$.
	Let $c_{i} = c(t_{i})$ denote the consumption at time $t_{i}$, $i = 1,2$. Let $(c_{1}, c_{2}) \mapsto u(c_{1}, c_{2}) \in \Rset$ denote an utility function that maps the consumption of different times to a scalar utility. Clearly, the ratio of $\frac{\partial u}{\partial c_{1}}$ and $\frac{\partial u}{\partial c_{2}}$ defines a discount factor, and hence an interest rate $\delta = \log(\frac{\partial u}{\partial c_{1}})-\log(\frac{\partial u}{\partial c_{2}})$, that relates the preference for a future value to the preference for a present value.

	A second component is the production functions that specifies the production of future goods from present time investments. Let $p$ denote a production function that maps time $t_{1}$ invested wealth to time $t_{2}$ consumption. Together with a function that translates spared time $t_{1}$ consumption to wealth, this gives another rate that translates time $t_{1}$ spared consumption to time $t_{2}$ consumption, not via preference, but via economic growth. Assuming a simple model, this rate is characterized by a product $\gamma \cdot g$, \cite{HepburnGosnell2014}.

	Together, the two parts define the \textit{growth adjusted social discount rate} $r = \delta + (\gamma-1) \cdot g$, where $\delta$ is sometimes called impatient rate or social time preference, $g$ describes the economic growth (relating today and future consumptions though production) and $\gamma$ describes the change in utility per change in consumption.

	Instead of a model for utility, consumption and production, we assume the existence of an effective financial market that already provides a (risky) discount factor, e.g., by the means of a bond market, and derive an - in a certain sense - risk-free discount factor from it.

	The direct specification of a financial market is compatible with a utility-consumption-production model since such a model would finally allow deriving the properties of the financial market.

	\subsubsection{Factoring-in Own Extinction}

	Another aspect that has been introduced in this context is the question if one should factor in the risk of extinction, see \cite{HepburnGosnell2014}. Some authors factor in the extinction of the human race to argue that the social discount rate is higher, namely by the intensity of the default probability $\lambda$ (assuming an exponential distribution). We explicitly mention this, because of two aspects:
	\begin{enumerate}
		\item The argument is similar to an effect that occurs in the financial industry, where a company's or bank's option to default leads to the curious effect that an increase of the default probability, that is, a decrease in the credit worthiness, leads to an increase of the value of the company. This effect is well know an it is comprised in the DVA and it is usually separated from the balance sheet, see above.

		\item Ironically, in our model, it is precisely this effect that leads to a \emph{decrease} of the discount rate, and possibly negative discount rates, because we factor in the need and cost to compensate the default probability. In a liquid financial marked, trading default protection one may determine a pure risk-free interest rate. In our setup we assume that default protection cannot be traded, hence it must be achieved by an approach that could require a certain over-compensation.
	\end{enumerate}
	The reason ``own extinction`` is not beneficial for the discount rate becomes transparent in our approach: default does not happen binary and on a global level, it occurs partially, and the remaining (surviving) parties have to compensate the liabilities of the defaulted party.
	
	\subsubsection{The Social Discount Rate may be Negative}

	Some authors consider a zero or even negative social discount rate to be implausible and provide arguments for a positive social discount rate, see~\cite{HepburnGosnell2014}. An argument provided is that a negative discount rate in a consumption model would imply that the economy is inefficient or require an unlimited sacrifice from present generations.
	
	We do not contradict this argument. Instead, we like to point out that this may be valid for the discount rate of an individual agent, a single asset in a risky financial market, or an infinitesimal or small disturbance of an economy in an equilibrium state, but it does not apply to large scale projects that need to be financed by multiple funding providers. In our model, the discount rate is allowed to depend on the notional through the need to secure sufficient default protection from multiple funding providers. See Section~\ref{sec:discountingDamage:socialDiscountRateMayBeNegative}.

	\subsubsection{Social Discounting and Long Term Rates}

	An obvious application for the model presented here is the valuation of \textit{damages} from climate change, which have to be repaired and are associated with high costs representing an extreme event.	For a discussion on the role of discounting to determine present values of future events related to damages from climate change, we refer the reader to~\cite{GollierWeitzman2010, Emmerling2019} and references therein.

	In~\cite{BroadieHugsthonSocialDiscounting, BiaginiGnoattoHaertelLongTermRates} the long maturity limit of interest rates is discussed, linking to the problem of valuation of long-term projects. It is stressed again there  that a major issue with discounting is that under certain assumptions, the discount factor is an exponential function of maturity, $\exp(-r (T-t))$, which results in strong underweight of future events.

	Let $P(T;t)$ denote the time $t$ value of a zero-coupon bond with maturity $T$ and $P(S,T;t)$ the time-$t$ value of a forward bond, that is, the time-$t$ value of  the price to be paid in $S$ to receive $1$ in $T$. The exponential discounting follows from the assumption of time-consistence ($P(S,T;t) = P(0,T-S;t) = P(T-S;t)$) and the absence of arbitrage.
	 via
	\begin{equation}
		\label{eq:discountingDamage:zeroBondReinvestment}
		P(T;t) = P(S;t) P(T-S;t) \text{.}
	\end{equation}
	However, the relation~\eqref{eq:discountingDamage:zeroBondReinvestment} assumes a re-investment strategy, that is, a trading strategy and neglects the possibility that the bonds used in the strategy defaults.

	\subsubsection{Valuation of a Financial Derivative where the Interest Rate depends on the Derivative Value}

	The idea that an interest rate may have a dependency on the value it is applied to has been introduced  in~\cite{EpsteinWilmottNewIRModel2011}. Here, instead of the classical bond valuation PDE\footnote{We used the notation from \cite{EpsteinWilmottNewIRModel2011}, where subscripts denote partial differentiation, $V$ the bond value, $\lambda$ the market price of risk, $r$ the interest rate and $K$ the bond coupon.}
	\begin{equation*}
		V_{t} + \frac{1}{2} \sigma V_{rr} + (u - \lambda \sigma) V_{r} - r V + K = 0
	\end{equation*}
	the authors consider the PDE
	\begin{equation*}
		V_{t} + f(r, V_{r}) V_{r} - r V = 0 \text{.}
	\end{equation*}
	While the authors neglect the diffusion $\frac{1}{2} \sigma V_{rr}$ (for simplification), the model considers a non-linear drift term, where $f$ depends on the sign of $V_{r}$. The authors establish this model to value a worst-case scenario, where borrowing and lending use different rates ($r^{+}$, $r^{-}$)  and rates are restricted by bounds. Note, that the PDE applies the dependency of the discount factor backward in time, depending on the future value of the derivative.

	\smallskip

	This situation is similar to the valuation of collateralized derivatives that became an active research topic 15 years later. For collateralized derivatives, the discounting is determined by a collateral contract. The collateral is given by the \textit{future value} of the derivative. A collateral contract may specify different rates for collateral posted and collateral received. This problem is related to the determination of the funding valuation adjustment (FVA) and has been studied extensively, see, e.g., \cite{BurgardKjaerPDE, BurgardKjaerBalance,FriesFundedReplication2011,LaurentAmzelekBonnaud2014} and references therein.

	\medskip

	The approach presented here is different from these situations as the dependency of the discount factor is propagated forward in time, not backward. The forward propagation is a result of funding providers defaulting, which impacts the interest that has to be paid for \emph{future} funding requirements.

	\subsection{Layout of the Paper}

	The main contribution of this paper is given in Section~\ref{sec:discountingDamage:discountingDamage} and~\ref{sec:discountingDamage:notionalDependecy} with a discussion of model properties in Section~\ref{sec:discountingDamage:implementation} to~\ref{sec:discountingDamage:applicationIAM}.

	In Section~\ref{sec:discountingDamage:reviewRiskNeutralValuation} we will shortly review discounting as it arises in the context of risk-neutral valuation. Risk-neutral valuation makes the assumption that claims can be replicated by trading in a market. Valuations are hence market-implied.

	In Section~\ref{sec:discountingDamage:discountingDamage} we discuss how funding for a future cash-flow may be provided when all market traded instruments are defaultable. Instead of bonds, we consider funding providers, that is, counterparties that can provide funding (that is, a zero-bond), but which are subject to default.
	The need to compensate for default by diversification introduces a discount factor that can be larger than 1 even if the market's interest rates are positive.

	In Section~\ref{sec:discountingDamage:notionalDependecy} we assume that the default probability of the funding provider depends on the required fund. This implies that the discount factor depends on the notional. That is, discounting is non-linear in the notional.

	In Section~\ref{sec:discountingDamage:implementation} we give a short discussion of the implementation and discuss some properties of the model in Section~\ref{sec:discountingDamage:propertiesOfTheModel}.

	We conclude in Section~\ref{sec:discountingDamage:numericalResults} by numerically investigating the properties of the model, mentioning an application to IAMs in Section~\ref{sec:discountingDamage:applicationIAM}.

	\clearpage

	\section{Risk Neutral Valuation and Market Implied Discounting}
	\label{sec:discountingDamage:reviewRiskNeutralValuation}

	Discounting as a time-value-of-money can be derived from a replication strategy, e.g., mapping future liabilities to current market prices, \cite{FriesLectureNotes2007, BrigoMercurio2007interest, FriesFundedReplication2011}.

	Consider a counterparty borrowing (unsecured) the amount $M$ from the market. The market requests an interest rate from the counterparty. This rate is called \textit{the funding rate}. Expressed as a continuously compounded rate, if the repayment of the borrowed  amount $M$ and all accrued interest occurs in $T$, then in $T$ the counterparty has to pay back the amount
	\begin{equation*}
		M \cdot \exp\left( \int_{0}^{T} r^{\mathrm{f}}(\tau) \mathrm{d} \tau \right)  \text{.}
	\end{equation*}
	Assuming a positive funding rate $ r^{\mathrm{f}}$, the amount paid back is larger than the amount $M$ originally borrowed.

	The funding rate $r^{\mathrm{f}}$ is often decomposed into two parts, $r^{\mathrm{f}} = r + \lambda$ . The rate $r$ is considered the risk-free rate, while $\lambda$ is a counterparty specific component reflecting the counterparty specific default risk.

	Hence, $r^{\mathrm{f}}$ is considered to be higher than an idealized \textit{risk-free rate $r$}, due to the perceived risk that the borrower can default, i.e., he can fail to pay back at the future time $T$.

	\medskip

	If the counterparty is a net borrower, i.e., at any future point in time it borrowed money from its investors, then any inflow of cash can be considered to earn the rate $r^{\mathrm{f}}$ by reducing the requirement to borrow money, hence reducing the funding costs.
	Under this situation, $N^{\mathrm{f}}(t) = \exp\left( \int_{0}^{t} r^{\mathrm{f}}(\tau) \mathrm{d} \tau \right)$ constitutes a numéraire for the counterparty (similar to a Bank account). Given that future values are stochastic the (risk-neutral) valuation of future cash-flows becomes the discounted expectation
	\begin{equation}
		\label{eq:discountingDamage:universalPricingWithFundingNumeraire}
		V(t) \ = \ E^{\mathbb{Q}}\left( V(T) \frac{N^{\mathrm{f}}(t)}{N^{\mathrm{f}}(T)} \ \vert \mathcal{F}_{t} \right) \ = \  E\left( V(T) \exp(-\int_{t}^{T} r^{\mathrm{f}}(\tau) \mathrm{d} \tau ) \ \vert \mathcal{F}_{t} \right) \text{,}
	\end{equation}
	where $N^{\mathrm{f}}(t)$ is the funding numéraire.

	This funding discounting \cite{FriesFundedReplication2011} can be understood from the assumption that the counterparty borrowed money from its investors and guaranteed the return $r^{\mathrm{f}}$ to them.

	In general, the funding $r^{f}$ is a stochastic process, and future cash-flows $V(T)$ are random variables.

	\subsection{Measures and Times}
	
	In~\eqref{eq:discountingDamage:universalPricingWithFundingNumeraire} the probability measure $\mathbb{Q}$ is induced by the assumption of risk-neutral replication.  That is, a risk-neutral valuation relies on the fact that contracts can secure future payments. Parameters derived from this context are \textit{market-implied parameters}. They reflect the market perceived (or market-implied) probabilities associated with the events. These parameters are a function of the time $t$ at which the contracts are traded.

	In contrast to the risk-neutral measure, the objective probability measure $\mathbb{P}$ of future events may differ from $\mathbb{Q}$ and parameters related to the real probability of events may differ from market-implied parameters.

	In the following, a parameter with a tilde denotes a parameter related to the objective probability measure, whereas the same symbol without the tilde denotes the corresponding market-implied parameter.
	
	\subsection{Rates and Compounding}
	
	The popular fundamental object for building interest rate curves is the zero-coupon bond: $P(T;t)$ is the time $t$ value of receiving $1$ in $T$. We may distinguish a default-free zero-coupon bond (denoted here by $P^{\circ}$) and a defaultable zero-coupon paying (denoted here by $P^{\mathrm{d}}$).

	Interest rates are an alternative (equivalent) form of expressing the system of zero bonds. Their compounding can be understood as a convention in their definition and is not necessarily related to a possible trading strategy.
	For example, we can express $P^{\circ}$ by a continuously compounding yield $r(T;t)$ or as a forward rate $L(t,T;t)$,
	\begin{equation*}
		r(T;t) \ = \ -\log(P^{\circ}(T;t))  / (T-t) \text{,} \qquad L(t,T;t) \ = \ \left( \frac{1}{P^{\circ}(T;t)}-1 \right) / (T-t) \text{.}
	\end{equation*}
	Similarly, a defaultable zero bond can be used to define an (implied)  survival probability, which is just
	\begin{equation*}
		\lambda(T;t) \ = \ -\log(P^{\mathrm{d}}(T;t))  / (T-t) - r \text{.}
	\end{equation*}

	In the following, we will often use the notation of continuously compounded rates, $\exp(-r(T;t) (T-t))$ and $\exp(-\lambda(T;t) (T-t))$, but this is only used because these expressions appear may be more familiar. Since the rates are time and maturity dependent, this does not imply an exponential decay.

	\subsection{Interest Rates}

	Concerning interest rates, we have to distinguish rates used to accrue collateral as part of collateralized products (EONIA, \euro-STR, SOFR) and rates of unsecured lending. For the latter, there are many. Any counterparty issuing bonds to receive funding creates a counterparty specific interest rate curve - the funding curve.

	Concerning the interest rate curve used to accrue collateral, these rates do not include the cost of providing funding since the daily settlement of collateralized contracts effectively removes counterparty risk.

	Hence, collateralization rates can be merely used to derive the information of the basis $r$ - an idealized risk-free rate. The rate is hypothetical because there is no long-term risk-free funding at this rate.

	In the following, we consider that funding is provided at the rate $r^{f}$ and use the risk-free rate $r$ as a basis to decompose the funding rate.

	\subsection{Survival Probabilities and Measures}
	
	The difference between the market-implied risk-neutral measure $\mathbb{Q}$ and the objective measure $\mathbb{P}$ becomes apparent if we consider default events.

	Assume that $V(T)$ is a deterministic time-$T$ value and $\indicatorfcn_{\tau < T}$ is the default indicator of $V(T)$. We assume that we can decompose its risk-neutral time-$t$ valuation as
	\begin{equation}
		\label{eq:discountingDamage:riskNeutralDefaultProb}
		E^{\mathbb{Q}}\left( V(T) \indicatorfcn_{\tau > T} \ \frac{N(t)}{N(T)} \ \vert \mathcal{F}_{t} \right) \ = \ V(T) \cdot \exp(-r (T-t)) \cdot \exp(-\lambda (T-t)) \text{,}
	\end{equation}
	where $r$ denotes some idealized averaged risk-free interest rate component (we use the same symbol $r$ as in $r(t)$ with a slight abuse of notation). Here $N$ is a risk free numéraire (as opposed to $N^{\mathrm{f}}$ above), and the default risk is encoded in the default indicator (and the filtration). The factor $\exp(-\lambda  (T-t))$ can be interpreted as the market-implied probability that the counterparty survives. That is, with probability $1-\exp(-\lambda (T-t))$ the cash-flow is not performed, so the cash-flow is in average
	\begin{equation*}
		V(T) \cdot \exp(-\lambda (T-t)) \ + \ 0 \cdot (1-\exp(-\lambda (T-t))) \ = \ V(T) \cdot \exp(-\lambda (T-t))
	\end{equation*}
	in $T$ and its risk-neutral valuation gives $V(T) \cdot \exp(-r (T-t)) \cdot \exp(-\lambda (T-t))$.

	\bigskip
	
	The view taken in equation \eqref{eq:discountingDamage:riskNeutralDefaultProb} is that of a market-implied survival probability, i.e., the term $\exp(-\lambda (T-t))$ is defined such that one matches observed market prices (reflecting the market view on the probability of default).
	
	In this situation, $\tilde{\lambda}$ denotes the corresponding parameter such that $\exp(-\tilde{\lambda} (T-t)) = E^{\mathbb{P}}\left( \indicatorfcn_{\tau > T} \ \vert \mathcal{F}_{t} \right) $ is the objective probability of survival of a funding provider from time $t$ up to time $T$. We will use this in the following Section~\ref{sec:discountingDamage:discountingDamage}.

	\clearpage
	\section{Compensating Default Risk by Diversification}
	\label{sec:discountingDamage:discountingDamage}

	The market-implied valuation of a liability is associated with the possibility that the liable counterparty may default on its liabilities. An increase in the default-probability reduces the present value of that liability from the market's perspective.

	In this section, we like to consider the valuation of cash-flows that have to be paid in all circumstances. That is, default is not an option.
	
	To ease notation, we will consider $t = 0$ and use the idealization from \eqref{eq:discountingDamage:riskNeutralDefaultProb}, that is, consider the three parts $\exp(-r T)$ (risk-neutral valuation), $\exp(- \lambda T)$ (implied survival probability) and $\exp(- \tilde{\lambda} T)$ (realized or objective survival probability). The discussion straightforwardly generalizes to the case of stochastic rates.

	\subsection{Buying Default Protection at a Market Price}
	
	In the risk-neutral valuation, we could try to cover the case of a non-performing counterpart. We seek protection for the default case having (market-implied) probability $1-\exp(-\lambda T)$, that is, the market price of this protection is $1-\exp(-\lambda T)$. Hence, buying protection the valuation becomes the risk-neutral valuation
	\begin{equation*}
		\underbrace{V(T) \exp(-r T) \exp(-\lambda T)}_{\text{value of defaultable cash-flow}} \ + \ \underbrace{V(T) \exp(-r T) (1-\exp(-\lambda T))}_{\text{\text{cost of defaultable protection}}} \ = \ V(T) \exp(-r T) \text{.}
	\end{equation*}
	
	\subsection{Guaranteeing a Payment from Diversified Funding}
	\label{sec:discountingDamage:discountingDamage:linear}

	The approach to value the liability by incorporating the market price of default protection depends strongly on the ability to buy that protection, on the business model that protection is actually bought, and on the reliability of the protection seller.

	The approach is possibly valid for liquid market products but unlikely feasible for catastrophic or systemic damages.

	\medskip
	
	An alternative approach to ensure the payment is to diversify the default risk. To start, consider an idealized setup and assume that we can contract payments with an objective survival probability $\exp(-\tilde{\lambda} T)$ from different counterparties and that their default events are independent. In that case, we can split a payment $X$ into $n$ parts across these counterparties and receive in expectation
	\begin{equation*}
		\sum_{i=1}^{n} \frac{1}{n} X \exp( -\tilde{\lambda} T ) \ = \ X \exp( -\tilde{\lambda} T) \text{.}
	\end{equation*}
	Assuming that the default events are independent, the variance of these repayments is given by
	\begin{equation*}
		\sum_{i=1}^{n} \frac{1}{n^{2}} X^{2} \exp( -\tilde{\lambda} T ) (1-\exp( -\tilde{\lambda} T )) \ = \ \frac{1}{n}  X^{2} \exp( -\tilde{\lambda} T) (1-\exp( -\tilde{\lambda} T )) \text{.}
	\end{equation*}
	
	Choosing $X = V(T) \exp(+\tilde{\lambda} T)$, we see that we receive $V(T)$ in expectation with a variance (risk) given by
	\begin{equation*}
		\frac{1}{n}  V(T)^{2}  (1-\exp( -\tilde{\lambda} T )) \text{.}
	\end{equation*}
	The risk can be reduced by increasing $n$.
	
	To summarize, we contract (distributed among multiple parties) the payment $X = V(T) \exp(+\tilde{\lambda} T)$. 
	
	\medskip
		
	A risk-neutral valuation of the future payment $X = V(T) \exp(+\tilde{\lambda} T)$ would give us
	\begin{equation*}
		V(T) \exp(- r T) \exp( (\tilde{\lambda} - \lambda) T)
	\end{equation*}
	and for $\tilde{\lambda} = \lambda$ we see the same value as for a discounting with the risk-free rate. 
	
	\medskip
	
	The discount factor $\exp(- r T) \exp( (\tilde{\lambda} - \lambda) T)$ contains three parts:
	\begin{itemize}
		\item $\exp(- r T)$ is a factor representing the (risk-neutral) time-value of money.
		
		\item $\exp(- \lambda T)$ is a discount we receive from a funding provider, due to its ability to default. This survival probability is fixed at trade time $t$ and is market-implied.
		
		\item $\exp(+ \tilde{\lambda} T)$ is the inverse of the true (objective) survival  probability and acts as a compensation of the (diversified) objective default risk. Note that it is observed at time $T$ if the funding is performed or not.
	\end{itemize}

	Reducing the risk by diversification, the default probability is observed at the future point in time $T$ under the objective probability measure, whereas the risk-neutral expectation is performed under the market-implied risk-neutral measure observed at the valuation time. In a stochastic model, we will be exposed to the risk of future changes in $\tilde{\lambda} - \lambda$.

	\subsection{Accounting for the Risk}
	\label{sec:discountingDamage:discountFactorFromDiversification}

	In Section~\ref{sec:discountingDamage:discountingDamage:linear}, diversification produces the required funding in expectation only. This is unsatisfactory and should be elaborated further.
	
	Assume that we distribute the total payment of $X^{*}$ among $n$ counterparties with i.i.d.~survival probabilities $\tilde{p}$, such that each entity pays $\frac{1}{n} X^{*}$ conditional to survival. 
	Let $Z$ denote the random variable representing the sum of the defaultable payments (a sum of independent Bernoulli distributed random variables). Then we receive in expectation .
	\begin{equation}
		\label{eq:discountingDamage:diversifiedExpectation}
		\mu \ = \ E(Z) \ = \ X^{*} \tilde{p} \text{.}
	\end{equation}
	The variance of the payment is
	\begin{equation}
		\label{eq:discountingDamage:diversifiedVariance}
		\sigma^{2} \ = \ V(Z) \ = \ \frac{1}{n} (X^{*})^{2} \tilde{p} (1-\tilde{p}) \text{.}
	\end{equation}

	For $n$ large, the random variable $Z$ can be approximated by a normal distribution and we can estimate the probability that the payment stays above a given threshold $\mu - c \sigma$ as
	\begin{equation*}
		P\left( Z \geq \mu - c \sigma \right) \ = \ 1-\alpha \ = \ \Phi(-c) \text{.}
	\end{equation*}
	For $\alpha = 1\%$ we find $c \approx 2.326$.\footnote{
	Alternatively, one might use the Cantelli inequality to estimate the probability that the payment stays above a given threshold. It is 
	\begin{equation*}
		P\left( Z \geq \mu - c \sigma \right) \ \geq \ 1 - 1 / (1+c^{2}) 
	\end{equation*}
	or $\alpha = 1 / (1+c^{2})$, i.e., $c = \sqrt{1/\alpha -1}$
	\begin{equation*}
		P( Z \leq \mu - \sigma \sqrt{\frac{1}{\alpha}-1} ) \leq \alpha \text{.}
	\end{equation*}
	For $\alpha = 1\%$ we have $c = \sqrt{99} = 9.9$. This is a much rougher estimate.
	}
	
	We now require that the amount $X = V(T)$ is paid with a given probability (confidence level) $1-\alpha$. Thus we require
	\begin{equation*}
		\mu - c \sigma \ = \ X \text{.}
	\end{equation*}
	Plugging in \eqref{eq:discountingDamage:diversifiedExpectation} and \eqref{eq:discountingDamage:diversifiedVariance}, i.e.~expressing $\mu$ and $\sigma$ in terms of the amount $X^{*}$ that has to be contracted, this gives
	\begin{equation*}
		X^{*} \tilde{p} - c \frac{1}{\sqrt{n}} X^{*} \sqrt{\tilde{p} (1-\tilde{p})} \ \stackrel{!}{=} \ X \text{,}
	\end{equation*}
	To ensure the payment of (at least) $X$ with a given probability $1-\alpha$ we thus have to contract the amount:
	\begin{equation}
		\label{eq:discountingDamage:discountFactorWithDiversification}
		X^{*} \ = \ X \frac{1}{\tilde{p} - c \frac{1}{\sqrt{n}} \sqrt{\tilde{p} (1-\tilde{p})}}
		\ = \ X \tilde{p}^{-1} \frac{1}{1 -  \frac{c}{\sqrt{n}} \sqrt{\tilde{p}^{-1}-1}} \text{.}
	\end{equation}
	With $\tilde{p} = \exp(- \lambda T)$ we now see that this gives
	\begin{equation*}
		X^{*} \ = \ X \ \exp(+ \tilde{\lambda} T) \ \frac{1}{1 -  \frac{c}{\sqrt{n}} \sqrt{\exp(+ \tilde{\lambda} T)-1}} \text{.}
	\end{equation*}
	
	The market price of these contracts (that is a risk-neutral valuation) would then give
	\begin{equation*}
		X \ \exp(- r T) \exp( (\tilde{\lambda} - \lambda) T) \ \frac{1}{1 -  \frac{c}{\sqrt{n}} \sqrt{\exp(+ \tilde{\lambda} T)-1}} \text{.}
	\end{equation*}

	We find that the need to diversify the funding risk modifies the discounting. The discount factor now consists of three parts:
	\begin{itemize}
		\item $\exp(- r T)$ is a factor representing the (risk-neutral) time-value of money.

		\item $\exp( (\tilde{\lambda} - \lambda) T)$, which is due to the fact that each funding supplier has a default risk and that there may be a mismatch between market-implied and realized default risk.
		
		\item The factor $\frac{1}{1 -  \frac{c}{\sqrt{n}} \sqrt{\exp(\tilde{\lambda}  T)-1}}$, which is due to the fact that we like to ensure the payments at a given confidence level via diversification among $n$ funding suppliers. Note that this factor is larger than $1$. For $n \rightarrow \infty$ the factor converges to $1$.
	\end{itemize}

	\bigskip

	A rough (first order) estimate for the additional factor is
	\begin{equation}
		\label{eq:discountingDamage:diversifiedRateAdjustment}
		\frac{1}{1 -  \frac{c}{\sqrt{n}} \sqrt{\exp(\tilde{\lambda} T)-1}} \ \approx \ 1 + \frac{c}{\sqrt{n}} \sqrt{\tilde{\lambda} T} \text{.}
	\end{equation}
	To translate the adjustment factor into an adjustment of the discount rate we can define the adjusted discount rate as
	\begin{equation*}
		r^{*} \ := \ r + (\lambda-\bar{\lambda}) + \log \left( 1 -  \frac{c}{\sqrt{n}} \sqrt{\exp(+ \tilde{\lambda} T)-1} \right) / T \text{.}
	\end{equation*}
	With $\lambda = \bar{\lambda}$ and $\log(1+x) \approx x$ and $\exp(x) \approx x$ we find
	\begin{equation}
		r^{*} \ \approx \ r - \frac{c}{\sqrt{n}} \sqrt{\lambda T} / T \text{.}
	\end{equation}

	If we consider only a limited amount of say $n=10$ funding suppliers, we find $c/\sqrt{n} \approx 3/4$, which indicates, that the factor can become a significant adjustment in the discounting, see Section~\ref{sec:discountingDamage:applicationIAM}.
	
	\clearpage
	\section{Notional-Dependent Discount Rate (Non-Linear Discounting)}
	\label{sec:discountingDamage:notionalDependecy}

	In the previous Section we considered $n$ independent funding providers for the given future funding requirement $X$.

	The fact that the discount factor depends on the true (objective) default probability motivates a further generalization: if we take the view of some (defaultable) funding provider, providing the amount $X$ (or a fixed fraction from it), it is reasonable that there is an upper bound to the fund that can be provided or - similarly - the objective (realized) default probability of the funding provider depends on the amount $X$. This assumption then introduces a notional dependency of the discount factor.
	
	Also, it is natural to consider the temporal distribution of funds provided by a funding provider. For example, if a funding provider provided the amount $X_{1}$ in $t_{1}$ and is required to provide the amount $X_{2}$ shortly after in $t_{2}$, then it is more likely that he defaults on $X_{2}$ if $X_{1}$ was high.

	\bigskip

	Both aspect are becoming relevant if we assume that there is a limited amount of funding providers, each having a limited capacity for providing funds at a certain survival probability.


	In other words, we like to consider two generalizations:
	\begin{itemize}
		\item we assume that an individual funding provider has a limited capacity, that is, it can provide funding (or put differently: pay for a damage) only within a certain limit.
		
		\item we assume that there is only a limited amount of funding providers.
	\end{itemize}
	If the ability to provide funding (or the funding rate) depends on the total amount of fund provided in the past, then this introduces an interdependence between different funding requirements, i.e. different cash-flows. This will make discounting a portfolio problem, similar to a CVA or MVA.
	
	Concerning the second assumption, one may argue that every individual could act as a funding provider, such that $n$ becomes the number of inhabitants, which makes $n$ large.	However, in that case, every funding provider can provide only a very limited amount.
	
	\medskip
	
	\subsection{Modelling a Notional Dependent Default Probability}

	A first idea to introduce a notional dependency would be to have a default intensity $\tilde{\lambda}$ dependent on the amount $X$ that has to be provided. It is reasonable to assume that the dependency is such that we default with higher probability only on the additional amount.
	
	Taking the default intensity to depend on the notional does not create a plausible model, because a funding requirement in $X_{1}$ in time $t_{1}$ should impact only future requirements in times $t_{2} > t_{1}$. This would then lead to inconsistent behaviour if a fund is spitted in two part, requested at almost the same time.

	Instead, we would like to have that the need to provide fund instantaneously affects the default probability of a funding provider on \emph{that} specific fund, that is if funding has to be provided in $T$, then the survival probability up to time $T^{-}$ (infinitesimal before $T$) is different from the survival probability up to time $T$. This effect cannot be captured by a classical model with a time-dependent intensity unless the intensity is allowed to become a Dirac distribution.

	Instead of starting with an intensity based model, we directly model a discontinuous change in the survival probability.\footnote{Since the intensity is a derivative of the survival probability, we see that the intensity has to become a Dirac distribution in that case.}

	\medskip

	At a future time $t$ the funding provider will default on a payment of $X$ with probability $1-\tilde{p}$, i.e., the expected fund provided in $t$ is $X \cdot \tilde{p}$. We assume that $\tilde{p}$ depends on the notional amount to be provided and that the ability to provide fund applies on a marginal basis, that is, for the expected fund provided $\tilde{X} = X \cdot \tilde{p}$ we have
	\begin{equation*}
		\mathrm{d} \tilde{X} \ = \ \tilde{q}(x) \mathrm{d}x \text{,} \qquad \tilde{X} (X=0) \ = \ 0 \text{,}
	\end{equation*}
	with some given monotone function $ \tilde{q}$. Furthermore, we assume that the marginal survival probability $\tilde{q}$ depends on the past fund provided, that is we assume that at time $t_{i}$ we have
	\begin{equation*}
		\mathrm{d} \tilde{X}_{i} \ = \ \tilde{q}(a(t_{i}) + x) \mathrm{d}x \text{.}
	\end{equation*}
	The term $a$ corresponds to the accumulated liabilities and models how the need to provide funds at previous times impacts the ability to provide fund at current times. A possible model for the funding consumption level $a$ is
	\begin{equation*}
		a(t_{i}) \ = \ \sum_{t_{k} < t_{i}} \tilde{X}_{k} \exp\left( - \alpha (t_{i} - t_{k}) \right) \text{.}
	\end{equation*}
	The parameter $\alpha$ represents some dampening, which could be justified by a growth of the funding provider; $\alpha$ just interpolates the limit cases $\alpha = 0$ and $\alpha = \infty$.

	\medskip

	With this model, we get a notional dependent effective survival probability $\tilde{p} = \tilde{p}(t_{i}, X)$ via
	\begin{equation*}
		 \tilde{p}(t_{i}, X) \ := \ \frac{1}{X} \int_{0}^{X}  \tilde{q}(a(t_{i}) + x) \mathrm{d}x \text{,}
	\end{equation*}
	that is
	\begin{equation*}
		\tilde{X} \ = \ X \cdot  \tilde{p}(t_{i}, X) \ = \ \int_{0}^{X}  \tilde{q}(a(t_{i}) + x) \mathrm{d}x \text{.}
	\end{equation*}

	In our numerical experiments, we choose a piece-wise constant, monotone decreasing survival probability function $\tilde{q}$.

	\bigskip

	\subsection{Analogy to Intensity-based Models with Poisson like Default Process}
	\label{sec:discountingDamage:notionalDependecy:intensityAsLimit}

	The modelling approach differs from that with a bounded default intensity $\lambda(t)$, where the survival probability is continuous in time $\exp(- \int_{0}^{t} \lambda(s) \mathrm{d}s )$. However, it is possible to establish a simple link between the two approaches - and this link will also help to understand the differences.

	Assume that the accumulated funding requirement $a(t) \ = \ \int_{0}^{t} \mathrm{d}\tilde{X}(s)$ is normal distributed, say $a(t)$ follows the stochastic differential equation
	\begin{equation*}
		\mathrm{d}a(t) = \mu \mathrm{d}t + \beta \mathrm{d}W(t),
	\end{equation*}
	i.e.~$a(t) = \mu t + \beta W(t)$. In this model, we have a linear increasing funding requirement and allow for some diffusion. The important aspect here is that the funding requirements are infinitesimal.

	Furthermore, assume that the (marginal) survival probability is an exponential function of the required fund, i.e., $\tilde{q}(x) = \exp(- x)$.
	Then we find for some incremental funding requirement $\Delta X(t)$
	\begin{equation}
		\label{eq:discountingDamage:linearizingForAnalogy}
		\Delta \tilde{X}(t) \ := \ \Delta X(t) \cdot  \tilde{p}(t, \Delta X(t)) \ = \ \int_{0}^{\Delta  X(t)}  \tilde{q}(a(t) + x) \mathrm{d}x
		\ \approx \ \Delta X(t) \tilde{q}(a(t)) \text{.}
	\end{equation}
	For $\Delta X(t)$ and $\tilde{q}(a(t))$ being independent, we find from
	\begin{equation*}
		\mathrm{E}\left( \tilde{q}(a(t)) \right) \ = \ \mathrm{E}\left( \exp(- \beta W(t)) \right) \ = \ \exp(- \lambda t)
	\end{equation*}
	with $\lambda = \mu - \frac{1}{2} \beta^2$ that
	\begin{equation}
		\label{eq:discountingDamage:intensityDiscountingAnalogy}
		\mathrm{E}\left( \Delta \tilde{X}(t) \right) \ \approx \ \mathrm{E}\left( X(t) \exp(- \lambda t) \right) \text{.}
	\end{equation}
	This last expression corresponds to a ``discounting'' with a survival probability $\exp(- \lambda t)$.
	
	While this is a straightforward construction, the funding requirements are mostly a linear function of time and translate state (x) to time (t), the analogy illustrates the difference to our approach: The analogy holds for small funding requirements $\Delta X(t)$, where past funding requirements are diffusive. Assuming the linearization used in \eqref{eq:discountingDamage:linearizingForAnalogy} it trivially creates independence of the discounted notional $X(t)$ and the discount factor.
	
	Hence, the fundamental difference in our approach is that we consider large notionals and their immediate effects on the survival probability.

	 \subsection{Notional dependent Discounting}

	In Section~\ref{sec:discountingDamage:discountFactorFromDiversification} we derived that to ensure the availability of funding of $X$ in $T$ within some confidence level $c$ we need to contract an amount $X^{*}$ that is chosen higher to compensate for the default. From our assumptions, we derived that $X^{*}$ is
	\begin{equation}
		\label{eq:discountingDamage:discountFactorWithDiversificationAgain}
		X^{*} \ = \ X \frac{1}{\tilde{p} - c \frac{1}{\sqrt{n}} \sqrt{\tilde{p} (1-\tilde{p})}}
		\ = \ X \tilde{p}^{-1} \frac{1}{1 -  \frac{c}{\sqrt{n}} \sqrt{\tilde{p}^{-1}-1}} \text{.}
	\end{equation}
	Here, $\tilde{p}$ denotes the objective survival probability (in contrast to the market-implied survival probability). 


	\medskip

	For the case that we are only interested in matching the funding requirement in expectation ($c=0$) the formula \eqref{eq:discountingDamage:discountFactorWithDiversificationAgain} simplifies to
	\begin{equation*}
		X^{*} \ = \ X \frac{1}{\tilde{p}} \text{.}
	\end{equation*}
	This has the simple interpretation that we need to contract $\frac{1}{\tilde{p}}$-times the original amount to compensate for a default of a funding-provider.

	Using this approach now on a marginal basis with the notional dependent survival probability this translates to the requirement
	\begin{equation}
		\label{eq:discountingDamage:relationXStartToX}
		\int_{0}^{X^{*}} \tilde{q}(a(t_{i}) + x) \mathrm{d}x \ \stackrel{!}{=} \ X \text{.}
	\end{equation}
	Put differently, with
	\begin{equation*}
		 \tilde{p}(t_{i}, X) \ := \ \frac{1}{X} \int_{0}^{X}  \tilde{q}(a(t_{i}) + x) \mathrm{d}x \text{,}
	\end{equation*}
	we have
	\begin{equation*}
		 X^{*} \tilde{p}(t_{i}, X^{*}) \ = \ X \text{.}
	\end{equation*}

	Here $\tilde{p}(t_{i}, X{*})$ is the effective survival probability for the amount $X^{*}$.
	In our applications, we usually know $X$ (the value that needs to be funded) and seek the corresponding factor $1 / \tilde{p}(t_{i}, X^{*})$. Thus, in our implementation, we are rather interested in the function
	\begin{equation*}
		 \tilde{p}^{*}(t_{i}, X) \ = \ \tilde{p}(t_{i}, X^{*})
	\end{equation*}
	where $X^{*}$ as a function of $X$ is given by \eqref{eq:discountingDamage:relationXStartToX}.

	\medskip

	In the following, we call $\tilde{p}(t_{i}, X)$ the survival probability (the expected percentage amount of $X$ achieve by contracting $X$) and $1/\tilde{p}^{*}(t_{i}, X)$ the default compensation factor, that is $1/\tilde{p}^{*}(t_{i}, X)-1$ is the percentage amount of $X$ required in addition to $X$ to ensure $X$ in expectation.

	\clearpage
 	\section{Implementation}
	\label{sec:discountingDamage:implementation}

 	For the implementation of the capacity of a funding provider, we need to implement a (stochastic) process that keeps track of the funding provided in the past (to calculate the level $a$) and provides the effective funding $\tilde{X} = X / \tilde{p}^{*}(t,X)$ for a funding request $X$.

	We consider a piece-wise constant $x \mapsto \tilde{q}(x)$ with $\tilde{q}(x) = \tilde{q}_{j}$ for $x_{j} < x < x_{j+1}$. Then the cumulated survival probability of the funding amount $X$, given a funding consumption $b$ is
	\begin{align*}
		\tilde{p}(t_{i}, b, X) & \ := \ \int_b^{b+X} \tilde{q}(\xi) \ \mathrm{d}\xi \\
		\intertext{with}
		\int_x^{y} \tilde{q}(\xi) \ \mathrm{d}\xi & \ = \
		\begin{cases}
			\tilde{q}_{l} (y-x) & \text{for $k > l$,} \\
			\sum\limits_{k \leq j < l} \tilde{q}_{j} (x_{j+1} - x_{j}) + \tilde{q}_{k} (x_{k}-x) + \tilde{q}_{l} (y-x_{l}) & \text{for $k \leq l$,}
		\end{cases}
	\end{align*}
	where $k = \min\{ j \ \vert \ x_{j} > x \}$ and $l = \max\{ j \ \vert \ x_{j} < y \}$.

	Likewise, we implement the function $\tilde{p}^{*}(t_{i}, b, X)$ that fulfils
	\begin{equation*}
		X \ = \ X^{*}  \tilde{p}(t_{i}, b, X^{*}) \text{,} \qquad \text{where} \qquad  X^{*} \ = \ X / \tilde{p}^{*}(t_{i}, b, X) \text{.}
	\end{equation*}
	Note that $X \mapsto X / \tilde{p}^{*}(t_{i}, b, X) = X^{*}$ is just the inverse of $ X^{*} \mapsto  X^{*}  \tilde{p}(t_{i}, b, X^{*}) = X$.

	If in $t_{i}$ a funding of $X_{i}$ is required, we  calculate the survival probability $\tilde{p}_{i}$ as
	\begin{equation*}
		\tilde{p}_{i}  \ = \ \tilde{p}(t_{i}, b(t_{i-1}), X_{i})) 
	\end{equation*}
	and the funding compensation as $1/\tilde{p}^{*}_{i} $ with
	\begin{equation*}
		\tilde{p}^{*}_{i}  \ = \ \tilde{p}^{*}(t_{i}, b(t_{i-1}), X_{i})) 
	\end{equation*}
	with
	\begin{align*}
		b(t_{-1}) & \ := \ 0 \\
		b(t_{i}) & \ = \ b(t_{i-1}) \exp(- \alpha (t_{i}-t_{i-1})) + X_{i} \text{.}
	\end{align*}
	Since $X_{i}$ is a random variable, the functions $\tilde{p}$ (survival probability), $\tilde{p}^{*}$ (default compensation factor), and $a$ (accumulated funding requirements) are random variables.

	\medskip

%
%
%

	The implementation of the functions can be found in~\cite{finmath-lib}, version 5.0.5.

	\clearpage
	\section{Properties of the Model}
	\label{sec:discountingDamage:propertiesOfTheModel}

	We summarize some properties of the model. We verify them in the numerical experiments in  Section~\ref{sec:discountingDamage:numericalResults}.

	\subsection{Portfolio Effects}

	Our model introduces a portfolio effect. The value of a portfolio of two products is different from the sum of the values of the two products, valued individually. Such a portfolio effect is also common in risk-neutral valuation, e.g., evaluating the cost of (netted) counterparty risk (CVA) or the cost of initial margin requirements (MVA).
	
	Since our model introduces a temporal dependency, where a survival probability for a funding depends on the accumulated past funding, we investigate financial products with periodic payments, see Section~\ref{sec:discountingDamage:results:temporalDependency}.
	
	\subsection{Non-Linearity, Variance Dependency}

	For classical linear products, like forward (rate) agreements or swaps, the volatility of the stochastic payments does not impact the valuation. This is due to the product valuation being linear and due to the existence of a static hedge.  This will be no longer the case in a model with a notional dependent survival probability since the default probabilities are state-dependent. In this case, scenarios with larger payments will obtain a different weight. This is (similar to) a wrong-way risk.
	
	To investigate the effect, we consider a forward rate agreement or a swap (having stochastic payments) and investigate the dependency on the interest rate volatility under our model. See Section~\ref{sec:discountingDamage:results:forward} and~\ref{sec:discountingDamage:results:asymmetry}.

	\subsection{Generation of a Continuum of Interest Rate Curves}

	Considering a defaultable interest rate swap with a notional dependent survival probability will introduce a continuum of par forward rates and par swap rates.

	Consider a stochastic interest rate term-structure model with stochastic forward rates
	\begin{equation*}
		L(T_{i}, T_{i+1}) = \frac{1}{T_{i+1}-T_{i}} \frac{P(T_{i})}{P(T_{i+1})}-1 \text{},
	\end{equation*}
	 where $P(T)$ is the stochastic process modelling the value of a zero-coupon bond with maturity $T$.

	In this model, consider the valuation of a swap, i.e., the stream of payments
	\begin{equation*}
		M \cdot \left( L(T_{i}, T_{i+1};T_{i}) - K \right) \qquad \text{in $T_{i+1}$ \quad ($i = 1,\ldots,n$)},
	\end{equation*}
	where $M$ denotes the notional. Performing a \textit{classical} valuation with respect to a numéraire $N$, the time-$t$ value of this stream of payments can be expressed as
	\begin{equation*}
		V_{\text{swap}}(M,K, T_{1},\ldots,T_{n+1};t) \ = \ M \sum_{i=1}^{n} \left( L(T_{i}, T_{i+1};t) - K \right) \cdot P(T_{i+1};t)  \text{,}
	\end{equation*}
	where
	\begin{align*}
		P(T_{i+1};t)  & \ = \ \mathrm{E}^{\mathbb{Q}^{N}}\left( \frac{N(t)}{N(T_{i+1})} \ \vert \ \mathcal{F}_{t} \right) \text{,} \\
		L(T_{i}, T_{i+1};t)  & \ = \ \mathrm{E}^{\mathbb{Q}^{N}}\left( L(T_{i}, T_{i+1};t) \frac{N(t)}{N(T_{i+1})} \ \vert \ \mathcal{F}_{t} \right) \big/ P(T_{i+1};t)  \text{.}
	\end{align*}
	The swap's par-rate $K^{*} = K^{*}(M, T_{1},\ldots,T_{n+1};t)$ is now defined as the rate for which
	\begin{equation*}
		V_{\text{swap}}(M,K^{*}, T_{1},\ldots,T_{n+1};t) = 0 \text{.}
	\end{equation*}
	Under our model, $K^{*}$ depends on the notional $M$. This dependence is in contrast to a classical valuation theory, where the par-rate does not depend on the notional $M$ and the classical forward rate for a single period is given by
	\begin{equation*}
		K^{*}(M, T_{i},T_{i+1};t) \ \stackrel{\text{classic model}}{=} \ L(T_{i}, T_{i+1};t) \text{.}
	\end{equation*}

	\medskip

	We now consider the par-rate of an unsecured swap with a notional dependent survival probability. This will result in the par rate to depend on the notional itself. The par-rate will receive a spread that depends on the slope (first derivative) of the survival probability as a function of the notional and the interest rate volatility of $L$.

	To understand this effect, let $X = L-K$ denote the stochastic cash-flow and $p = p(X) = p(0) + p^{\prime}(0) X$ a notional dependent survival probability with $p(0)$, $p^{\prime}(0)$ being deterministic. Then we have
	\begin{equation*}
		X p(X) \ = \ X p(0)+ X p^{\prime}(0) X \ = \  X p(0)+ X^{2} p^{\prime}(0) \ = \ X p(0) + X^{2} \frac{\partial \log(p(X))}{\partial X}\big\vert_{X=0} p(0)
	\end{equation*}
	Now, if $K^{\circ}$ is such that for $X^{\circ} = L-K^{\circ}$ we have $\mathrm{E}( X^{\circ} p(0) ) = 0$, then we find for $K = K^{\circ} + \Delta K$
	\begin{equation*}
		0 \ \stackrel{!}{=} \ \mathrm{E}\left( X p(X) \right) \ = \ -\Delta K P(T_{i+1};t) p(0) + \mathrm{E}\left( X^{2} \frac{\partial \log(p(x))}{\partial x}\big\vert_{x=0} p(0) \right)
	\end{equation*}
	that
	\begin{equation*}
		\Delta K \ = \ \frac{1}{P(T_{i+1};t)} \mathrm{E}\left( X^{2} \right) \frac{\partial \log(p(x))}{\partial x}\big\vert_{x=0} \text{.}
	\end{equation*}
	In this equation, the right-hand side has a (weak) dependence on $\Delta K$, due to $X = X^{\circ} - \Delta K$. Instead of solving for $\Delta K$, it is - for obtaining the right intuition - sufficient to approximating $X$ by $X^{\circ}$ and see that
	\begin{equation}
		\label{eq:discountingDamage:spreadAsFunctionOfVol}
		\Delta K \ \approx \ \frac{1}{P(T_{i+1};t)} \mathrm{E}\left( (X^{\circ})^{2} \right) \frac{\partial \log(p(x))}{\partial x}\big\vert_{x=0} \text{.}
	\end{equation}
	This is the impact of the notional dependent survival probability on the par rate: The term $\mathrm{E}\left( (X^{\circ})^{2} \right) $ is the variance of the underlying cash-flow, the term $\frac{\partial \log(p(x))}{\partial x}\big\vert_{x=0}$ is the slope of the log-survival probability.

	Note that $\frac{\partial \log(p(x)}{\partial x}\big\vert_{x=0}$ is negative if the survival probability decreases for increasing (positive) notional and that in this case, the spread $\Delta K$ is negative. We verify this behaviour in our numerical experiments in Section~\ref{sec:discountingDamage:results:generationInterestRateCurve}.

	\newpage
 	\section{Numerical Experiments}
	\label{sec:discountingDamage:numericalResults}

	\subsection{Setup}

	\subsubsection{Models}
	
	We use classical models to simulate the evolution of (market) risk factors, based on Itô processes, like a Black-Scholes model, Bachelier model or LIBOR market model, \cite{BrigoMercurio2007interest, FriesLectureNotes2007}.

	We use these models to simulate the  ``funding requirements'' as pay-offs of classical financial derivatives (like a swap or a forward contract or forward rate agreement). The simulation and valuation is performed under a risk-neutral measure.

	We chose this setup for the sake of comparison. Since the valuation of these products is well known under these modes, we can investigate the impact of a notional dependent discounting. For $\lambda = \bar{\lambda}$, $c = 0$, $\alpha = \infty$ we recover the classical risk-neutral valuation of a defaultable cash-flow.

	The parameter $\bar{\lambda}$ is the objective future default intensity, while $\lambda$ is the market-implied default intensity (a discount on defaultable loans). A reasonable approach could be to set $\lambda = 0$ and just consider some excess default intensity $\bar{\lambda}$.
	
	The parameter $c$ defines the quantile level of risk we are willing to allow for funding mismatches. If $c = 0$ then funding is provided only in expectation.
	
	The parameter $n$ specifies the number of independent funding providers and impacts the risk to miss the required funding. For $c = 0$ the parameter $n$ has not significance (see~\eqref{eq:discountingDamage:discountFactorWithDiversificationAgain}).

	The parameter $\alpha$ controls how fast the funding system recovers.

	In our experiments, it is sufficient to specify the funding rate $r^{f} = r + \lambda$ (or, consider $\lambda = 0$). All cash-flows are considered defaultable. In that case $\tilde{\lambda}$ can be interpreted as the mismatch of the realized (objective) default intensity and the market-implied default intensity.

	\subsubsection{Probability Measure}

	If payments are stochastic, risk-neutral valuation values future scenarios by taking their expectation under a risk-neutral measure. This approach is justified by the ability to replicate payments by trading activities. If replication is not possible, the scenarios should be simulated under the objective measure, and instead of expectation,  risk measures (like expected shortfall) should be considered.
	
	That said, we conduct our analysis under the risk-neutral measure since this allows for an analytic benchmark (in some cases) and a change of measure would not impact the qualitative behaviour. 

	\subsection{Numerical Results}

	We present some numerical results, illustrating the behaviour of the model. The experiments in this sections are available in~\cite{finmath-experiments}, using~\cite{finmath-lib}. Source code is provided through the referenced repositories.

	\subsubsection{Analogy to Intensity Based Models}
	
	As illustrated, a state-dependent default probability may translate to a time-dependent survival probability, in expectation, in a simplified model, where the funding requirements are infinitesimal, and distributed over time.

	We consider a process $\mathrm{d}S = \mu \mathrm{d}t + \sigma \mathrm{d}W$.  We use an exponential, state-dependent, instantaneous survival probability $q(x) = \exp(-x)$.

	This implies (see~\eqref{eq:discountingDamage:intensityDiscountingAnalogy}) a stochastic survival probability for the interval from $0$ to $T$ being
	\begin{equation*}
		\exp(-\lambda(0,T) T) \qquad \text{with} \quad \lambda = \mu - 0.5 \sigma^{2} \text{.}
	\end{equation*}

\begin{figure}[!ht]
		\centering
		\includegraphics[width=0.9\linewidth]{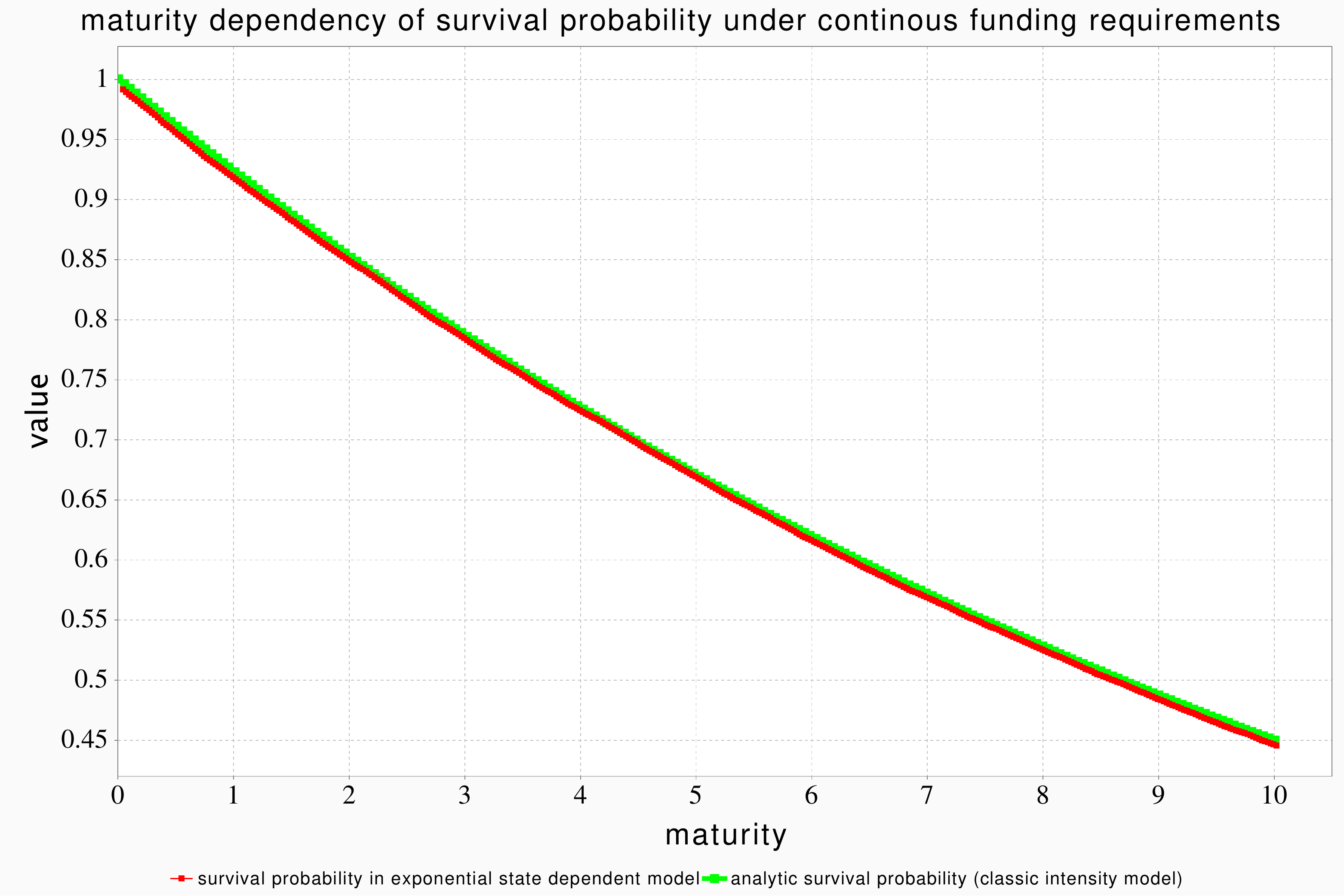}
		\caption[
		]{
			Maturity dependence of the (expectation of) the stochastic survival probability in a model of state-dependent instantaneous funding requirements (red) and the benchmark limit case of a deterministic exponential survival probability (green). The slight visible difference is a result of the time-discretization used in the model of the stochastic survival probability.
		}
		\label{fig:maturity-dependency-of-survival-probability-in-continuous-funding}
		\addtocounter{cpfNumberOfFigures}{1}
	\end{figure}

	Figure~\ref{fig:maturity-dependency-of-survival-probability-in-continuous-funding} depicts the numerical result for $\mu = 0.1$, $\sigma = 0.2$ using a model with a state-dependent default probability and a classical discounting with the default intensity $\lambda$.
	Verifying the relation, the result can be seen as basic test of the implementation of the state-dependent model - at least for this limit case.

	\subsubsection{A Forward Contract with Non-Linear Discounting}
	\label{sec:discountingDamage:results:forward}

	We consider the value $X(T)$ to be log-normal distributed following a Black-Scholes model $\mathrm{d}X = r X \mathrm{d} + \sigma X \mathrm{d}W(t)$. We assume that $X(T)$ represents a future cash-flow requirement (e.g. the cost to compensate damage).

	A risk-neutral valuation of $X(T)$ would result in $X(0)$, independent of the parameter $\sigma$. If $X(T)$ is considered to be a defaultable cash-flow, where the default is assumed to be independent of $X$, we will arrive at a value
	$X(0) \exp(- \lambda T)$. Following our discussion in Section~\ref{sec:discountingDamage:discountingDamage}, this would imply that we need to contract the amount
	$X(0) \exp( (\tilde{\lambda} - \lambda) T)$ to compensate for the default - at least, in expectation.

	Considering a notional-dependent default probability, we consider the default-compensated amount $X(T) / \tilde{p}^{*}(T, 0, X(T))$. This is the amount we have to diversify among defaultable funding providers to receive $X(T)$ in expectation.
	If $\tilde{p}^{*}$ is neither a constant nor homogenous, this will introduce a non-linearity, and hence a risk-neutral valuation will depend on the volatility $\sigma$.

\begin{figure}[!ht]
		\centering
		\includegraphics[width=0.9\linewidth]{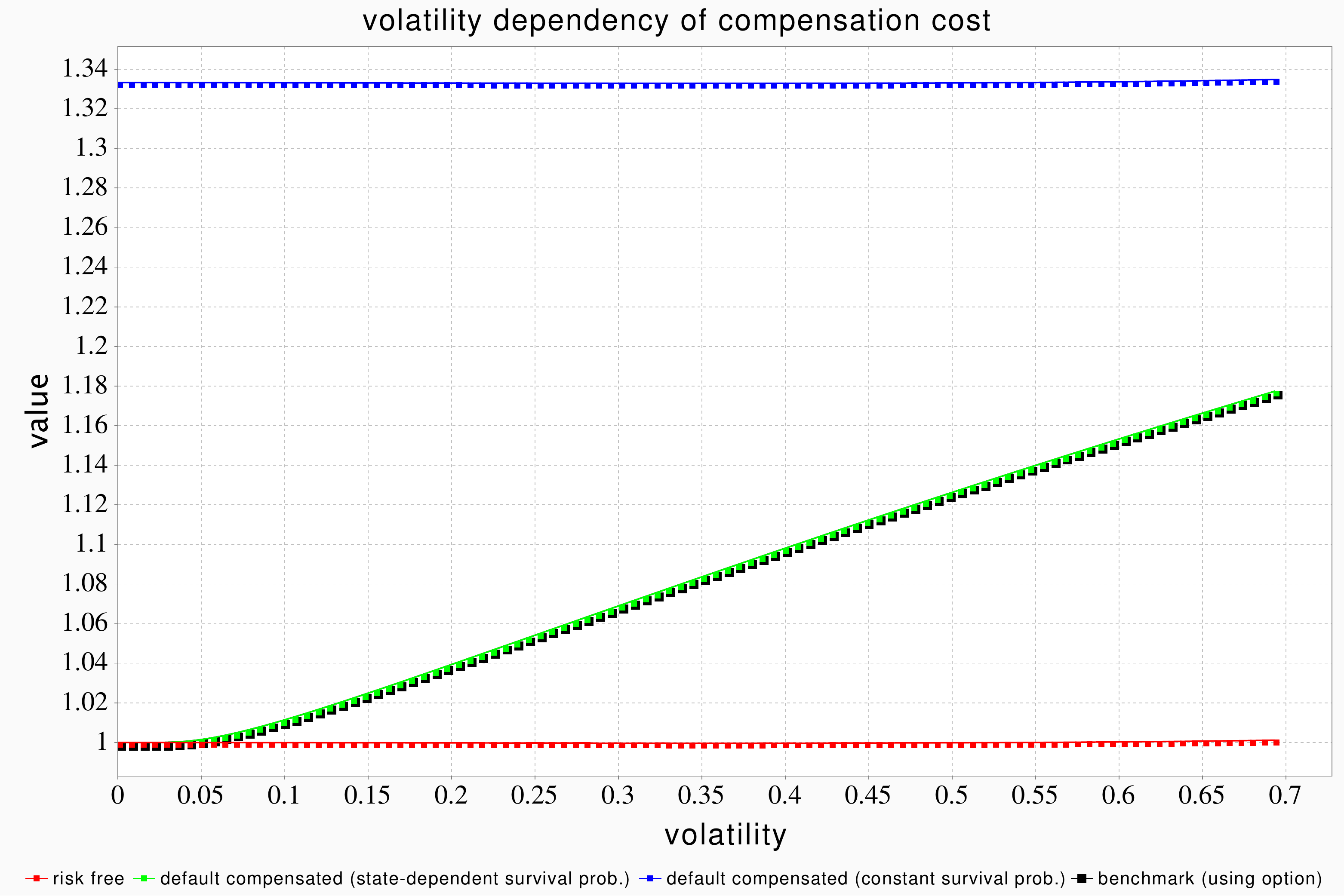}
		\caption[
		]{
			Volatility dependency of the valuation of $X(T)$, including the state-dependent default-compensation cost (green), without default-compensation cost (red) and the valuation corresponding to a constant instantaneous default intensity.
		}
		\label{fig:volatility-dependency-of-compensation-cost-x}
		\addtocounter{cpfNumberOfFigures}{1}
	\end{figure}

	If we consider a simple model with a piece-wise linear function $p$ generated from a piece-wise constant function $q$, the function $x \mapsto x \ \tilde{p}^{*}(T, 0, x)$ will be piece-wise constant too. We take
	\begin{equation*}
		\tilde{p}^{*}(T, 0, x) \ = \ \begin{cases}
			1.0 & \text{for $x < L$} \\
			a & \text{for $x > L$.}
		\end{cases}
	\end{equation*}
	Then $X(T) / \tilde{p}^{*}(T, 0, X(T))$ corresponds to a pay-off of a European option. It is
	\begin{equation*}
		X(T) / \tilde{p}^{*}(T, 0, X(T)) \ = \ X(T) + (1 - 1/a) \max(X(T)-L,0) \text{.}
	\end{equation*}
	From this, the volatility dependence of the pay-off becomes obvious. Figure~\ref{fig:volatility-dependency-of-compensation-cost-x} depicts the situation for $T=5$, $a=0.75$.

	An obvious upper bound to the risk-neutral valuation of the default-compensate pay-offs is $(1 - 1/a) X(0)$.

	\subsubsection{Asymmetry}
	\label{sec:discountingDamage:results:asymmetry}

\begin{figure}[!ht]
		\centering
		\includegraphics[width=0.9\linewidth]{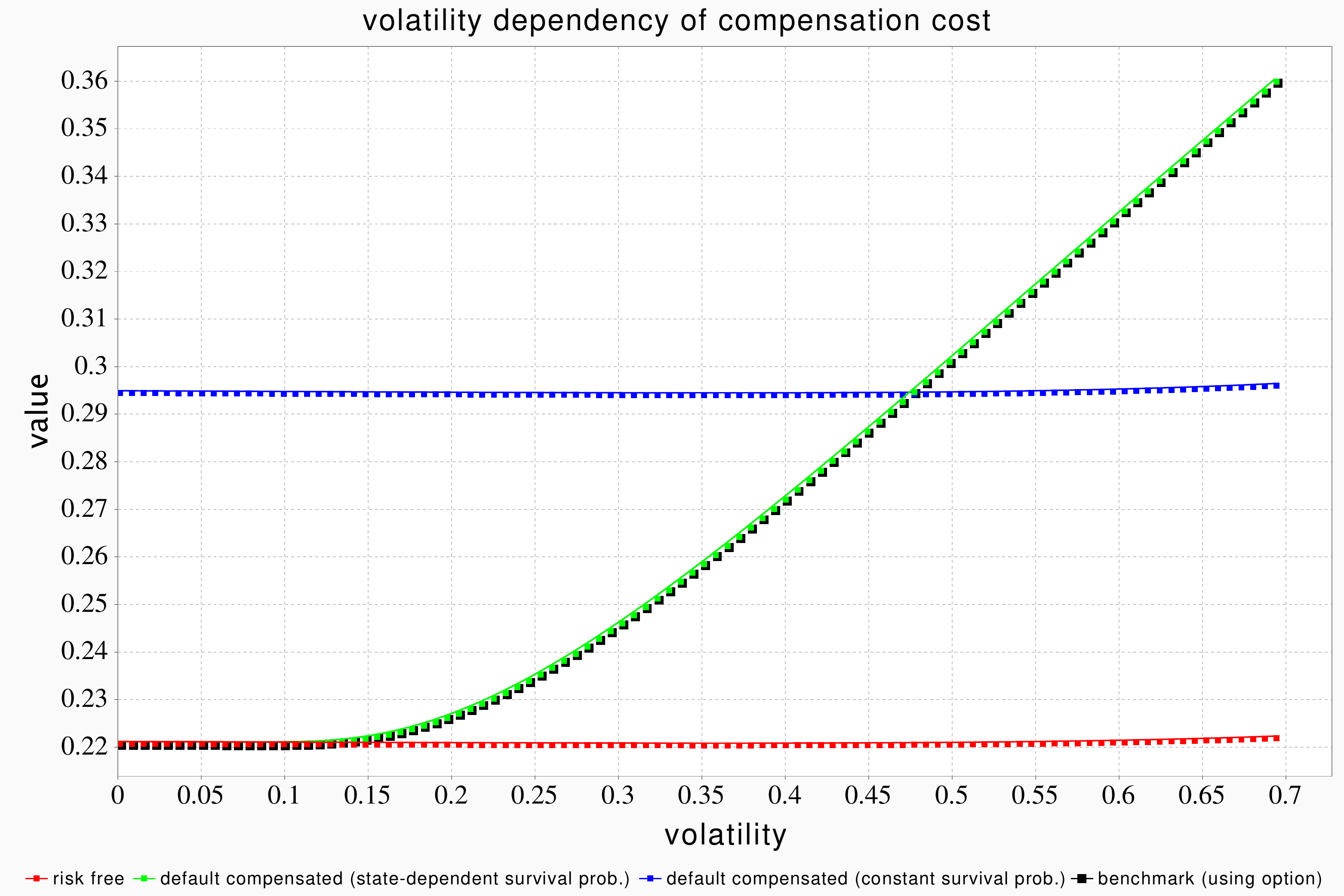}
		\caption[
		]{
			Volatility dependency of the valuation of$X(T)-K$, including the state-dependent default-compensation cost (green), without default-compensation cost (red) and the valuation corresponding to a constant instantaneous default intensity.

			The setup corresponds to that in Figure~\ref{fig:volatility-dependency-of-compensation-cost-x}, excepts that the cash-flows allow for negative values (here interpreted as gains).
		}
		\label{fig:volatility-dependency-of-compensation-cost-x-k}
		\addtocounter{cpfNumberOfFigures}{1}
	\end{figure}

	In the example of Section~\ref{sec:discountingDamage:results:forward} the impact of the non-linear discounting appears as a simple, almost linear interpolation between the two extreme factors $1$ and $1/p_{0}$. The situation looks different if we consider a future cash-flow requirement of $X(T)-K$, that is, a forward agreement with forward value $K$. We assume a log-normal $X(T)$ as above. The amount $X(T)-K$ can be positive or negative, where we consider the positive value a damage (liabilities) and the negative value a gain.\footnote{We consider damages to be potentially unbounded, but gains bounded.} We used a non-linear discounting, i.e., default-compensation that compensates only the positive amounts. In relative terms the effect then appears much stronger and it is not bound by applying the factor $1/p_{0}$ to the all paths, see~\ref{fig:volatility-dependency-of-compensation-cost-x-k}.

	\subsubsection{Temporal Dependency}
	\label{sec:discountingDamage:results:temporalDependency}

	We consider a sequence of values (damages)
	\begin{equation}
		\label{eq:discountingDamage:cashflowStream}
		X(T_{i})-K\text{,} \qquad \text{for $i = 1,\ldots,n$.}
	\end{equation}

	If the non-linear discounting model is applied to the cash-flow stream \eqref{eq:discountingDamage:cashflowStream}, the compensation factor (and the survival probability) of the value $T_{i}$ depends on the events $T_{j}$, $j < i$. In addition, since there is a positive correlation between $X(T_{i})$ and $X(T_{j})$ being high, there is a feedback effect. In case of a non-linear discount factor, we have that the sum of the individual compensated values is different from the compensated sequence of values.
	
\begin{figure}[!ht]
		\centering
		\includegraphics[width=0.9\linewidth]{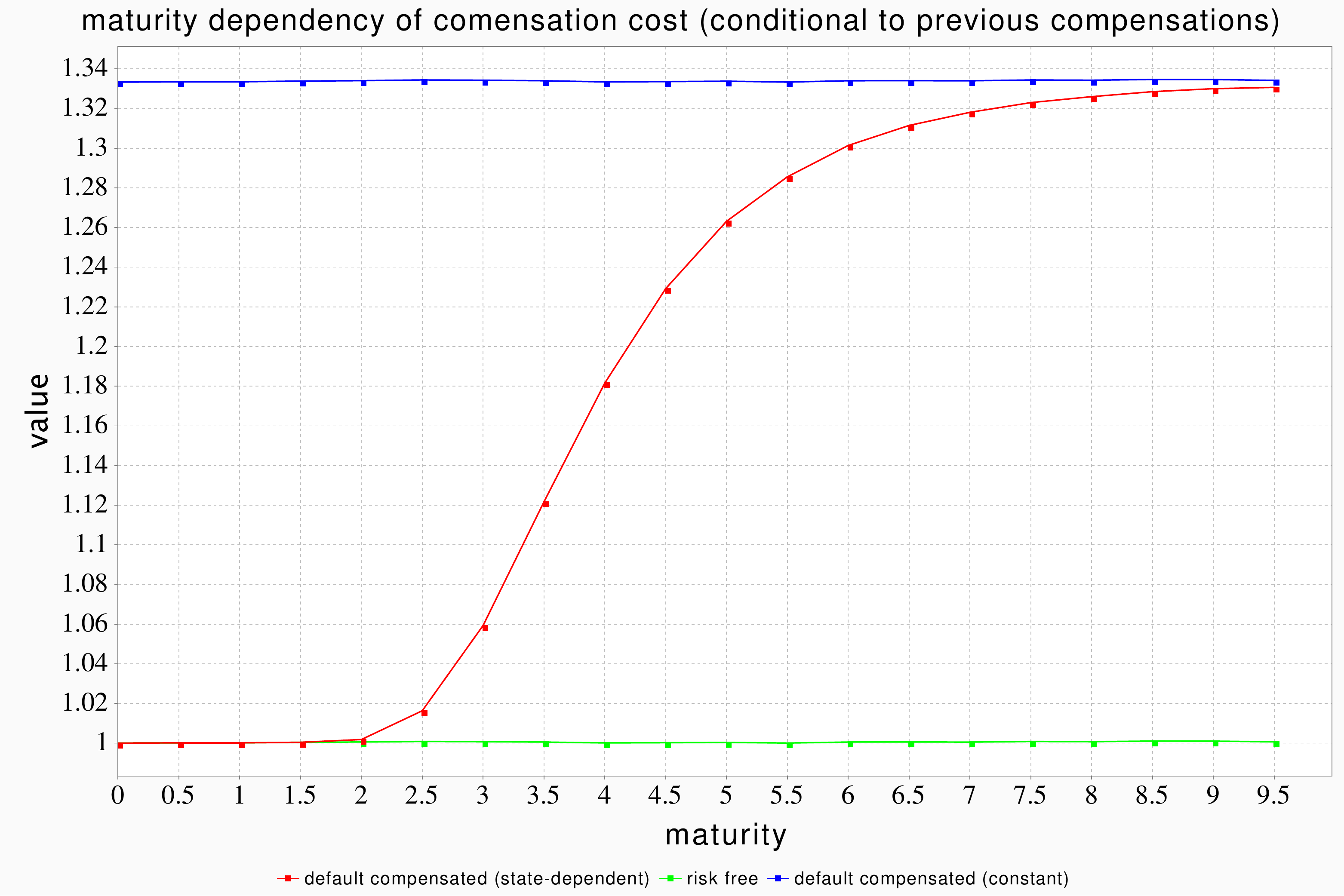}
		\caption[
		]{
			Maturity dependence of the valuation including (stochastic) default compensation cost (red) of a sequence of funding requirements $X(T_{i})$, ($i = 1,\ldots,n$).
		}
		\label{fig:maturity-dependency-of-compensation-cost-x-0-0}
		\addtocounter{cpfNumberOfFigures}{1}
	\end{figure}

\begin{figure}[!ht]
		\centering
		\includegraphics[width=0.9\linewidth]{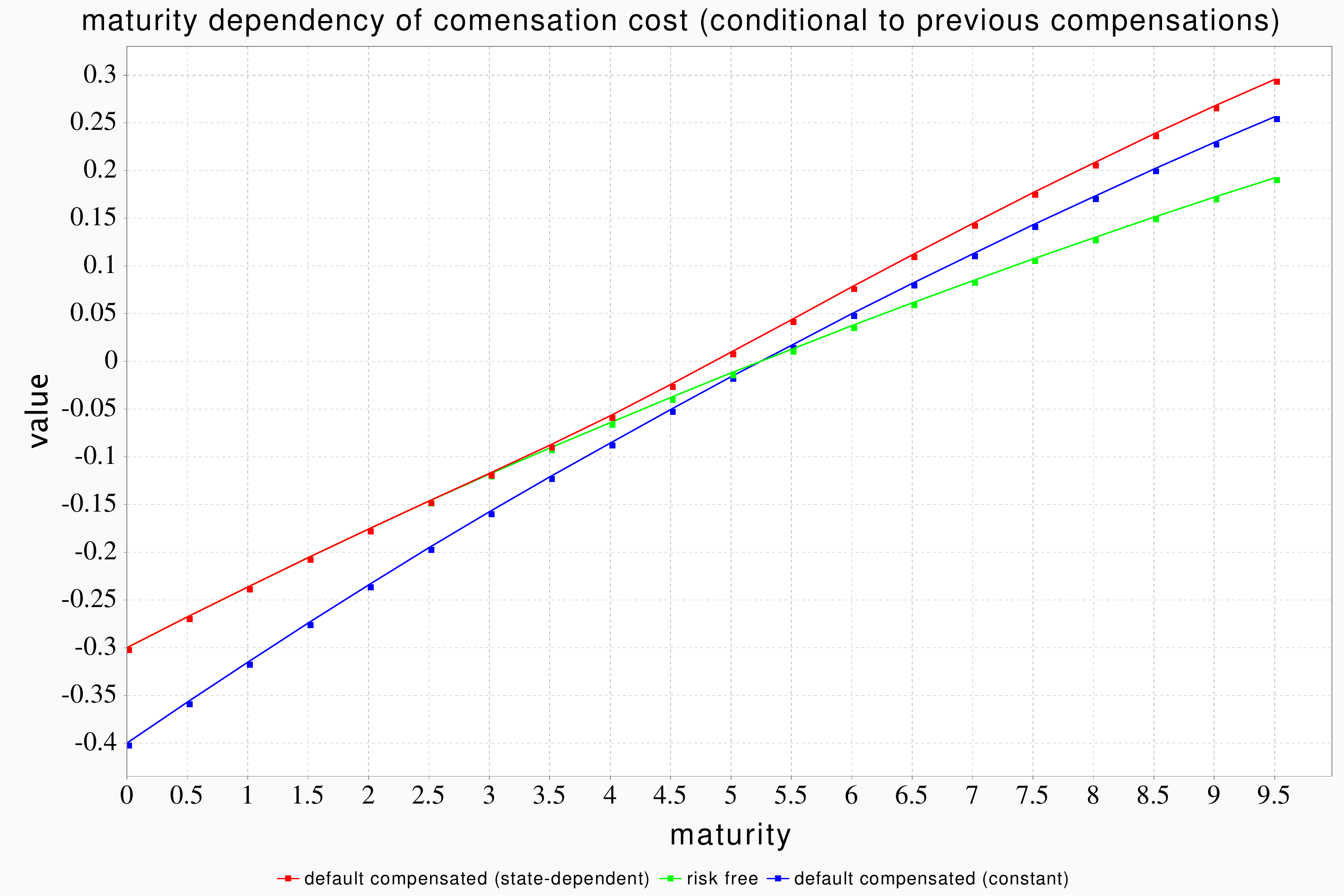}
		\caption[
		]{
			Maturity dependence of the valuation including (stochastic) default compensation cost (red) of a sequence of funding requirements $X(T_{i})-K$, ($i = 1,\ldots,n$).
		}
		\label{fig:maturity-dependency-of-compensation-cost-x-1-3}
		\addtocounter{cpfNumberOfFigures}{1}
	\end{figure}

	In Figure~\ref{fig:maturity-dependency-of-compensation-cost-x-0-0} and~\ref{fig:maturity-dependency-of-compensation-cost-x-1-3}, we depict the valuation of the single amount $X(T_{i})-K$ as a function of $T_{i}$, conditional to the prior compensation of the amounts $X(T_{j})-K$ for $j < i$. Figure~\ref{fig:maturity-dependency-of-compensation-cost-x-0-0} shows the result for $K=0$, Figure~\ref{fig:maturity-dependency-of-compensation-cost-x-1-3} for $K > 0$.

	Since in a classical (risk-free) setup the valuation of $X(T_{i})-K$ is independent of  prior valuations of $X(T_{j})-K$ and the risk-free valuation of any of those is $X(0)-K P(T_{i})$, for the risk-free valuation we will see a horizontal line for $K = 0$ and an upward sloping curve for $K > 0$. Likewise, a constant compensation factor will result in a parallel shift of the line corresponding to the risk-free valuation.

	A state-dependent survival probability and hence a state-dependent compensation leads to a strong maturity dependency. For the case $K=0$ the value is just a maturity dependent interpolation between the two constant cases.

	For $K > 0$ the behaviour in our test is as follow: for low maturities almost all scenarios result in negative values (which we interpret as gains) and the compensation factor is (path-wise) $1$, given the same result as the risk-free case. For higher maturities more scenarios show positive values (positive funding requirements), consuming the capacity of the funding provider, decreasing the marginal survival probability. In that case, the funding compensation can exceed that of a constant compensation factor, due to the asymmetry between positive and negative values.

	%

	\clearpage

	\subsubsection{Generation of a Continuum of Interest Rate Curves}
	\label{sec:discountingDamage:results:generationInterestRateCurve}

	We numerically verify the result that a notional dependent survival probability generates a continuum of (defaultable) par-rates (swap rates and forward rates).

	The model used is a standard forward rate model (LIBOR market model) with an exponentially decaying forward rate volatility. The initial forward rate curve is flat (at $5\%$).

	The notional dependent survival probability was $1.0$ if the notional stayed below a certain threshold, this leads to the induced interest rate spread being (almost) zero for small notionals.

\begin{figure}[!ht]
		\centering
		\includegraphics[width=0.9\linewidth]{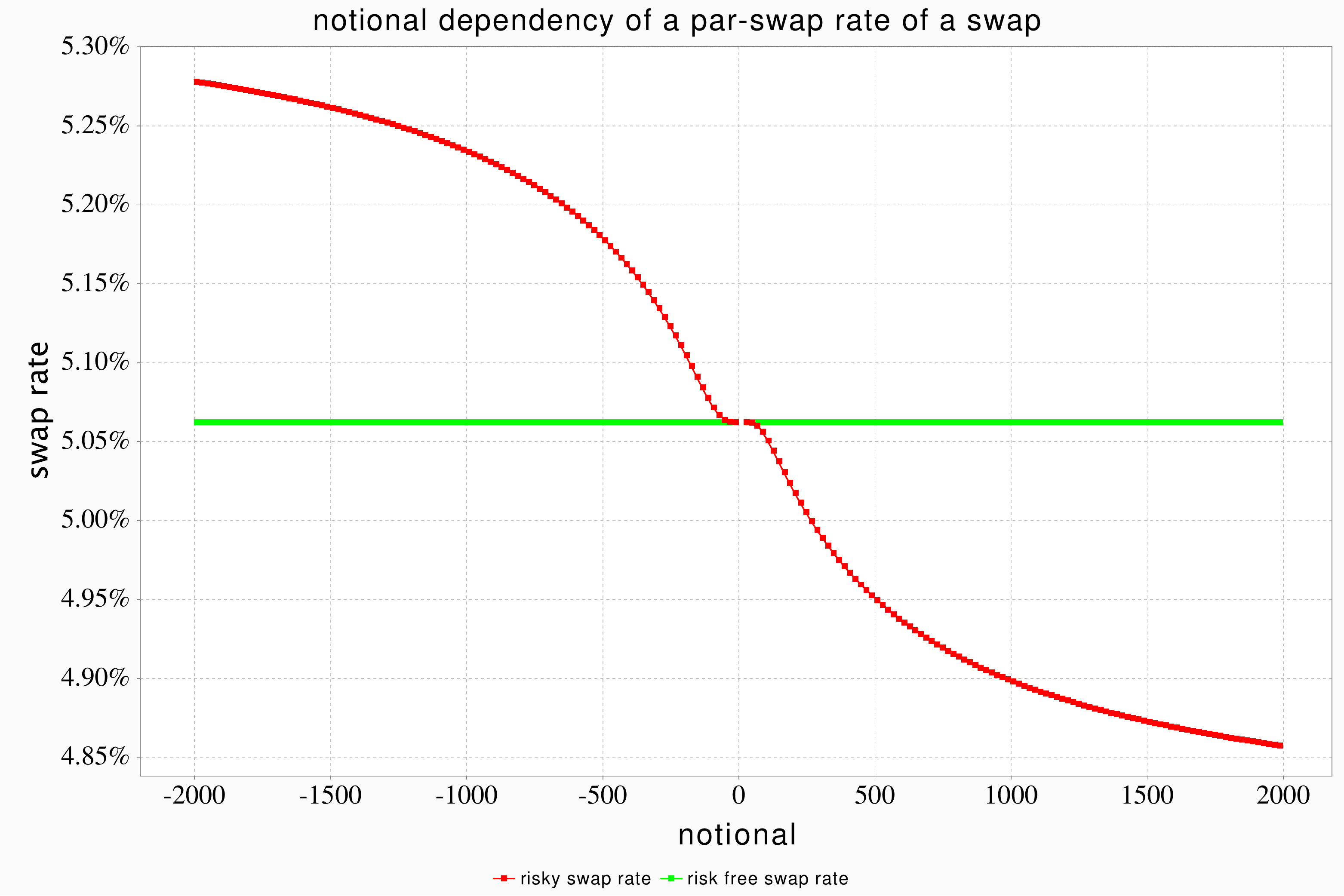}
		\caption[
		]{
			The par swap rate of a 20Y swap with a notional dependent survival probability (red). In our model, the survival probability is $1$ (i.e., default-free) as long as the cash-flow is below a fixed threshold. Hence the swap rate coincides with the risk-free swap rate (green) for small notional. A large notional generates a spread.
		}
		\label{fig:notional-dependency-of-par-swap-rate-with-survival-probability}
		\addtocounter{cpfNumberOfFigures}{1}
	\end{figure}

	Figure~\ref{fig:notional-dependency-of-par-swap-rate-with-survival-probability} shows the dependency of the 20Y par swap rate on the notional. 

\begin{figure}[!ht]
		\centering
		\includegraphics[width=0.90\linewidth]{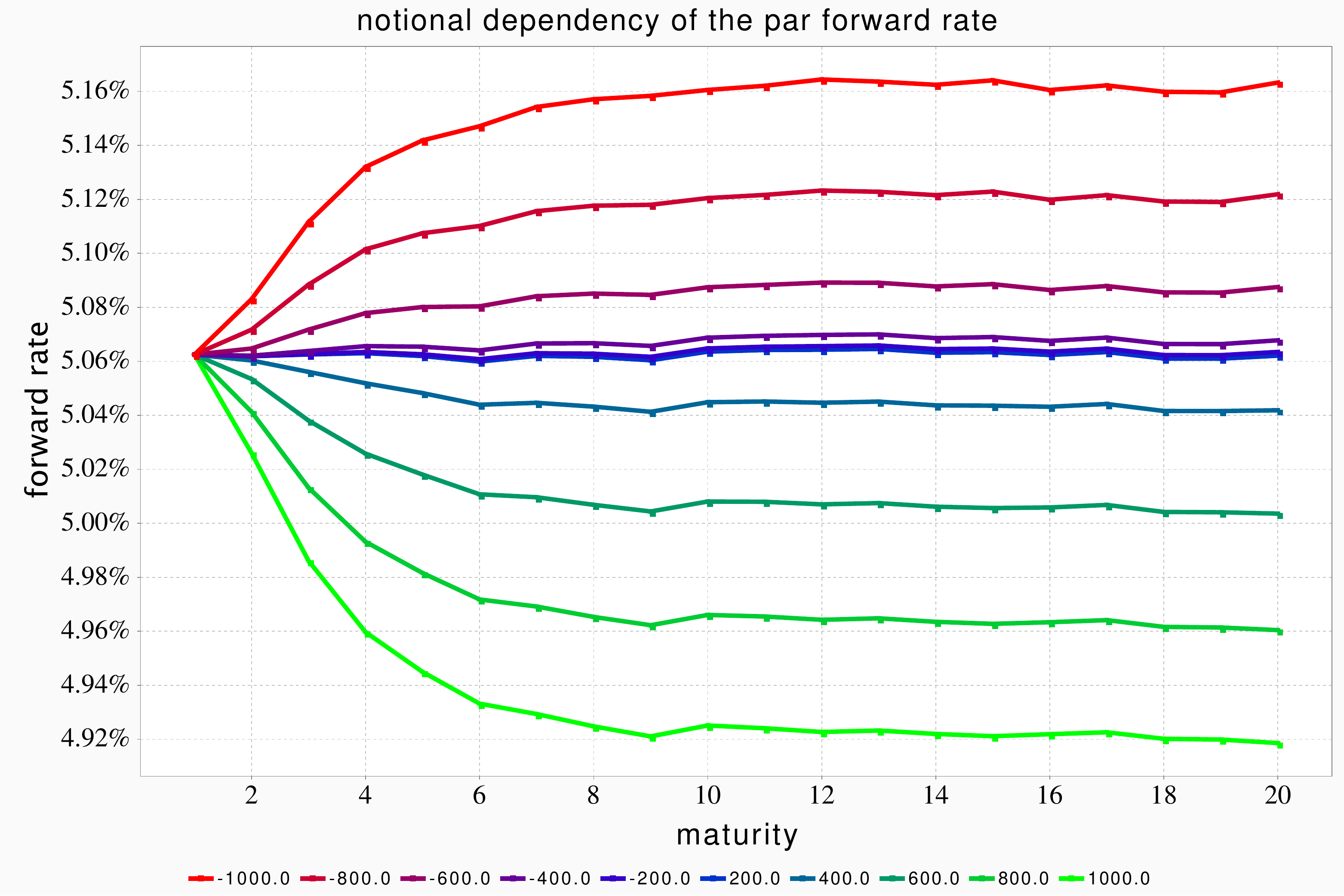}
		\caption[
		]{
			The forward rate curve, i.e., the curve of the par rate of a forward rate agreement, for different notionals. The shape of the curves is a consequence of an exponential decay in the forward rate volatility as a function of time-to-maturity. While this leads to a mean reversion of the short rate, it also leads to a flattening of the spread curve, as higher maturity rates have approximately the same volatility.

			The slightly non-smooth shape of the curve is because they are obtained via a Monte-Carlo simulation of the model.
		}
		\label{fig:notional-dependency-of-forward-rate-curve-with-survival-probability-volatility-0.5}
		\addtocounter{cpfNumberOfFigures}{1}
	\end{figure}

\begin{figure}[!ht]
		\centering
		\includegraphics[width=0.9\linewidth]{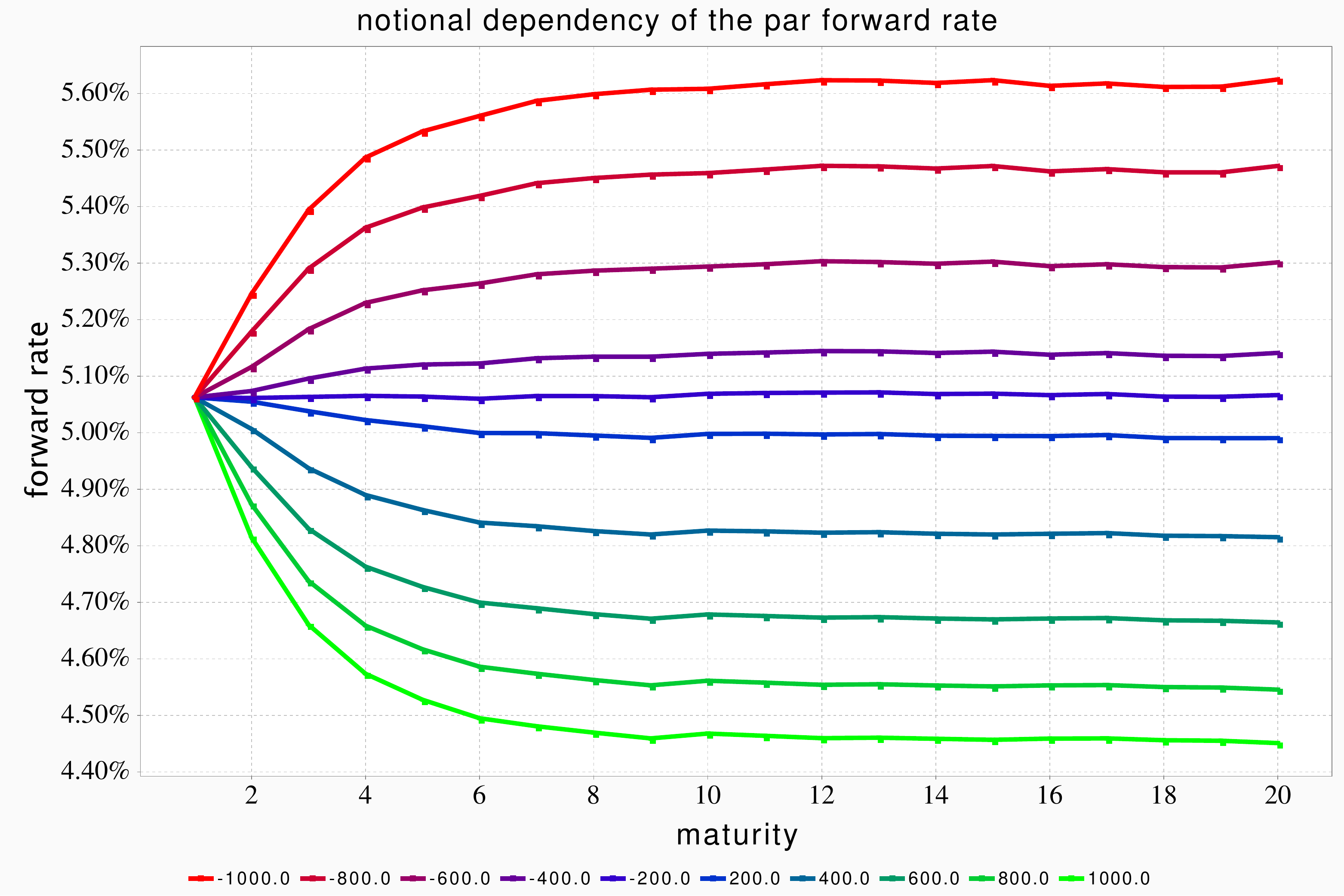}
		\caption[
		]{
			The forward rate curve for different notionals from the same experiment as in Figure~\ref{fig:notional-dependency-of-forward-rate-curve-with-survival-probability-volatility-0.5}, except for the volatility being twice as large. This results in an increase of the spread by (approximately) a factor of four, corresponding to our derivation~\eqref{eq:discountingDamage:spreadAsFunctionOfVol}.
		}
		\label{fig:notional-dependency-of-forward-rate-curve-with-survival-probability-volatility-1.0}
		\addtocounter{cpfNumberOfFigures}{1}
	\end{figure}

	In Figure~\ref{fig:notional-dependency-of-forward-rate-curve-with-survival-probability-volatility-0.5} and~\ref{fig:notional-dependency-of-forward-rate-curve-with-survival-probability-volatility-1.0} the same model is used to calculate the (par) forward rate curve for different notionals of the forward rate agreement and different volatilities of the interest rate, verifying the intuition derived above.
	
	\clearpage

	\section{Application to Integrated Assessment Models of Climate Change Impact}
	\label{sec:discountingDamage:applicationIAM}

	Integrated assessment models (IAM) try to combine geo-physical properties (like emissions and atmospheric temperature) with economic quantities (damages and abatement costs). The valuation includes, of course, a discounting.

	The resulting \textit{social costs of carbon} have a strong dependency on the discount rate used.

	A simple IAM is DICE, \cite{Nordhaus1518}, which already exhibits the strong dependency on the discount rate. For example, Figure 3 in \cite{Nordhaus1518} shows that the social cost of carbon using a rate of $1\%$ is approximately four times higher than that using a rate of $3\%$.

	\subsection{The Social Discount Rate of an IAM may be Negative}
	\label{sec:discountingDamage:socialDiscountRateMayBeNegative}

	In Section~4 of this paper we proposed a notional dependent discount rate.
	Our numerical experiments illustrate that such a model results in notional dependent interest rate curves. Here, an average market level may be a positive interest rate, while in extreme scenarios we have negative interest rates.

	In Section~3 of this paper we suggested a possible model for a discount rate that establishes a notional dependency by considering a default probability of funding providers to depend on the size of the funded amount.

	If we think of damages resulting from climate change creating such extreme notional, we may just consider equation~\eqref{eq:discountingDamage:diversifiedRateAdjustment}.

	If we take comparably mild parameters, e.g.~a maturity of $T = 25$ years, a growth adjusted social interested rate $r = 1\%$ (that does not account for the effect of non-linear discounting), taking $n = 10$ funding providers with an individual default probability of $\lambda = 1\%$ and confidence level of $1\%$ we end up at a social interest rate, adjusted for non-linear discounting,
	\begin{equation}
		r^{*} \ \approx \ r - \frac{c}{\sqrt{n}} \sqrt{\lambda T} / T \ \approx \ 1\% - \frac{3}{4} \sqrt{0.01 / 25} = -0.5 \% \text{.}
	\end{equation}

	\paragraph{Future Research}
	
	In~\cite{FiesQuante2022} we combine the framework with an integrated assessment model and study the effect of various non-linearities.
	
	\clearpage

	\section{Conclusion}

	We derived a model for \textit{discounting}, i.e., the valuation - or the assessment - of a future liability in terms of  an equivalent present value (or cost). We considered a default compensation factor ($\exp(+\tilde{\lambda} T)$) that takes the role of the cost to buy default protection, for the case where default protection is not available as a traded asset. Not accounting for tail risk, i.e., achieving protection only in expectation, the factor is the inverse of the objective survival probability. Factoring in the risk to fail in providing sufficient default protection, the factor increases, depending on the number of funding providers and their default probabilities.

	We then established a model where the objective survival probability is not an exponential function of time, but a function of the fund required. This leads to a non-linear discount factor - that is - the valuation becomes a non-linear function of the notional.

	We investigated the properties of the model by comparing the valuation of liabilities under classical models. It was shown that the model create a continuum of interest rate curves.

	We provided a prototypical  open source implementation of the framework.

	Our approach is not contradicting the existing theory, it should be rather seen as an extension, introducing an additional non-linear effect. In fact, we saw from Section~\ref{sec:discountingDamage:notionalDependecy:intensityAsLimit} that the classic intensity based model is a limit case for infinitesimal small funding requirements.

	\clearpage

	\printbibliography

	\newpage

\section*{Notes}

\subsection*{Suggested Citation}

\begin{itemize}
	\item[] \sloppypar \textsc{Fries, Christian P.}: Non-Linear Discounting and Default Compensation (April, 2020/21). \newline
	\url{https://ssrn.com/abstract=3650355}
	\newline
	\url{http://www.christian-fries.de/finmath}
\end{itemize}

\subsection*{Classification}

\begin{small}
	
	\noindent Classification:
	\href{http://www.ams.org/msc/}{MSC-class}: 91G30 
	\\
	\phantom{Classification:}
	\href{http://www.aeaweb.org/journal/jel_class_system.html}{JEL-class}: Q51, H43, C63.
	\\


	
%

	
	
	\noindent Keywords:
	Valuation,
	Discounting,
	Interest Rate Modelling,
	Funding,
	Environmental Damage
	
	
	
	\vfill
	
	\bigskip
	\centerline{\small\thepage \ pages. \thecpfNumberOfFigures \ figures. \thecpfNumberOfTables \ tables.}
	
\end{small}

\end{document}